\documentclass[sigconf,nonacm]{acmart}
\AtBeginDocument{%
  }

\setcopyright{acmlicensed}
\copyrightyear{2018}
\acmYear{2018}
\acmDOI{XXXXXXX.XXXXXXX}
\acmConference[Conference acronym 'XX]{Make sure to enter the correct
  conference title from your rights confirmation email}{June 03--05,
  2018}{Woodstock, NY}
\acmISBN{978-1-4503-XXXX-X/2018/06}

\usepackage{xcolor,colortbl}
\usepackage[ruled,vlined]{algorithm2e}
\SetKwInput{KwInput}{Input}
\SetKwInput{KwOutput}{Output}
\usepackage[most]{tcolorbox}
\usepackage{tabularx}
\usepackage{newfloat}
\usepackage{listings}
\usepackage{soul}
\usepackage{color} 
\usepackage{caption} 
\usepackage{multirow}
\usepackage{multicol}
\usepackage{url}
\usepackage{longtable}

\newtcolorbox{boxEnv}[1][]{
    center,
    left=0mm,
    top=0.25mm,
    right=0mm,
    bottom=0.25mm,
    colframe=gray!90!black,
    colback=black!5!white, 
    boxrule=0.5pt,
    title=#1
}

\newcommand{\fname}{AgenticCyOps\xspace}
\newcommand{\facronym}{Securing Multi-Agentic AI Integration in Enterprise Cyber Operations\xspace}

\begin{document}

\title[\fname]{\fname: Securing Multi-Agentic AI Integration \\in Enterprise Cyber Operations}

\renewcommand{\authorsaddresses}{}
\author{Shaswata Mitra, Raj Patel, Sudip Mittal, 
Md Rayhanur Rahman, Shahram Rahimi}
\affiliation{%
  \institution{Department of Computer Science, 
  The University of Alabama, Tuscaloosa, USA}
  \country{}}
\email{{smitra3, rpatel38, sudip.mittal, mrahman87, shahram.rahimi}@ua.edu}

\renewcommand{\shortauthors}{Mitra et al.}

\begin{abstract}

    Multi-agent systems (MAS) powered by LLMs promise adaptive, reasoning-driven enterprise workflows, yet granting agents autonomous control over tools, memory, and communication introduces attack surfaces absent from deterministic pipelines. While current research largely addresses prompt-level exploits and narrow individual vectors, it lacks a holistic architectural model for enterprise-grade security. We introduce \fname (\textit{\facronym}), a framework built on a systematic decomposition of attack surfaces across component, coordination, and protocol layers, revealing that documented vectors consistently trace back to two integration surfaces: tool orchestration and memory management. Building on this observation, we formalize these integration surfaces as primary trust boundaries and define five defensive principles: authorized interfaces, capability scoping, verified execution, memory integrity \& synchronization, and access-controlled data isolation; each aligned with established compliance standards (NIST, ISO 27001, GDPR, EU AI Act). We apply the framework to a Security Operations Center (SOC) workflow, adopting the Model Context Protocol (MCP) as the structural basis, with phase-scoped agents, consensus validation loops, and per-organization memory boundaries. Coverage analysis, attack path tracing, and trust boundary assessment confirm that the design addresses the documented attack vectors with defense-in-depth, intercepts three of four representative attack chains within the first two steps, and reduces exploitable trust boundaries by a \textit{minimum} of 72\% compared to a flat MAS, positioning \fname as a foundation for securing enterprise-grade integration.
\end{abstract}

\keywords{Agentic AI, Multi-Agent Systems (MAS), Cyber Defense, Cybersecurity Operations, Security Operation Center (SOC), SOAR}


\maketitle

\section{Introduction}\label{section:introduction}

    The pattern of speculative over-investment followed by gradual, transformative real-world impact has characterized every major technological revolution of the past two centuries~\cite{perez2002technological}. With the theoretical groundwork for AI and agentic systems established in the late 20th century and recent advances in generative AI enabling autonomous planning, tool invocation, and natural language coordination, enterprise adoption is rapidly shifting from deterministic, rule-based agent pipelines to autonomous orchestration~\cite{gartner2025agentic}. However, history warns that rapid paradigm adoption without adequate security carries systemic risk: web proliferation enabled injection attacks that dominated OWASP Top 10 rankings over a decade~\cite{halfond2006classification}, cloud migration introduced misconfiguration vulnerabilities exposing over 100 million records~\cite{khan2022systematic}, and IoT deployment with default credentials enabled botnet weaponization of 600,000 devices~\cite{antonakakis2017understanding}.

    Multi-agent systems (MAS) powered by LLMs now face an analogous inflection~\cite{tomavsev2026intelligent}: agents that autonomously invoke tools can be redirected to malicious endpoints~\cite{triedman2025multi, narajala2025enterprise}; shared memory that enables collective intelligence equally facilitates privacy breaches~\cite{chen2024blockagents, wei2025memguard}; and emergent collusion can arise without explicit adversarial prompting~\cite{agrawal2025evaluating, mathew2025hidden}. The security research landscape addressing these threats remains fragmented, with proposals primarily targeting prompt-injection exploits and narrow individual vectors. While adversarial robustness of individual LLMs remains an active concern, MAS introduces a distinct threat class: attacks that exploit the \textit{integration architecture} connecting agents to tools and shared state, independent of any single model's vulnerability. Moreover, existing authorization standards such as OAuth~2.1 were not designed for autonomous, long-running agent sessions~\cite{owasp2025agentic}. Yet the field still lacks a unified threat model that maps MAS vulnerabilities to actionable defensive mechanisms. Specifically, three questions remain open: \textit{how do MAS attack vectors map to exploitable architectural surfaces}, \textit{what defensive principles can systematically address them}, and \textit{whether such principles can be instantiated in high-stake domains with measurable coverage}.

    To address the first, we conduct a systematic decomposition of MAS attack vectors across three abstraction layers: component, coordination, and protocol, analyzing how each vector exploits the underlying architecture. This decomposition reveals a consistent structural pattern: despite their diversity, documented vectors converge on two exploitable integration surfaces, \textit{tool orchestration} and \textit{memory management}.

    For the second, we treat these surfaces as primary trust boundaries and derive five defensive design principles: three governing tool orchestration (authorized interfaces, capability scoping, verified execution) and two governing memory management (integrity \& synchronization, access control with data isolation), each principle reusing established security primitives~\cite{narajala2025securing, kim2025prompt, zou2025blocka2a, wei2025memguard, mao2025agentsafe} and aligned with compliance standards including NIST SP~800-207, ISO~27001, GDPR, and the EU AI Act~\cite{nist_cybersecurity_framework, euaiact2024}.

    For the third, we select Cybersecurity Operations (CyberOps) as a proving ground. Unlike other enterprise domains where integration failures result in operational disruption, CyberOps is \textit{inherently adversarial}: agents routinely process attacker-crafted artifacts through security tools and accumulate threat intelligence that adversaries are directly incentivized to corrupt, placing both integration surfaces under simultaneous, intentional pressure. Moreover, a compromised CyberOps agent does not merely malfunction; it actively shields the adversary from the infrastructure designed to detect them. The operational urgency reinforces this choice: attackers now traverse networks in under 30 minutes~\cite{lyngaas2024crowdstrike}, yet SOC teams still require a reported mean of 181 days to identify a breach and an additional 60 days to contain it~\cite{totalassure2025detection}, an asymmetry compounded by alert fatigue and knowledge fragmentation across disconnected tools~\cite{assaf2025socburnout}. We apply the framework to a Security Orchestration, Automation, and Response (SOAR) architecture adopting the Model Context Protocol (MCP)~\cite{mcp_architecture_2025} as its structural basis, embedding security as an architectural constraint rather than a runtime policy overlay, and evaluate the design through coverage analysis, attack path tracing, and trust boundary assessment. This paper makes the following contributions:

    \begin{itemize}
        \item \textbf{Attack Surface Decomposition.} A systematic analysis of MAS threats across component, coordination, and protocol layers, identifying that documented vectors converge on tool and memory integration surfaces (Table~\ref{tab:attack_surface_mapping}).
        
        \item \textbf{Defensive Design Framework.} Five security principles derived from the decomposition, where each documented vector is addressed by at least two complementary principles grounded in compliance mandates (Table~\ref{tab:defense_coverage}).
        
        \item \textbf{CyberOps Application and Evaluation.} An agentic SOAR architecture with phase-scoped agents, consensus-validated execution, and organizational memory, evaluated through coverage analysis (Table~\ref{tab:coverage_matrix}), attack path tracing (Table~\ref{tab:attack_paths}), and trust boundary assessment (Table~\ref{tab:trust_boundaries}) demonstrating a \textit{minimum} of 72\% reduction in exploitable boundaries.
    \end{itemize}

    \noindent We organize the paper using a \textit{What--Why--How--Next} progression: agentic AI fundamentals and attack surfaces (\S\ref{section:agenticai}), defensive principles (\S\ref{section:design_principles}), MAS-integrated CyberOps framework with coverage, attack path, and trust boundary evaluation (\S\ref{section:agentic_cyber_tools}), and trade-offs with open challenges (\S\ref{section:discussion}), before concluding. The appendix provides a list of acronyms, a categorized summary of the literature underpinning our research foundation, and full boundary enumeration.

\section{Agentic AI and Attack Surfaces (\textit{The What})}\label{section:agenticai}
    
    \begin{figure*}[!ht]
        \centering \includegraphics[width=1\textwidth]{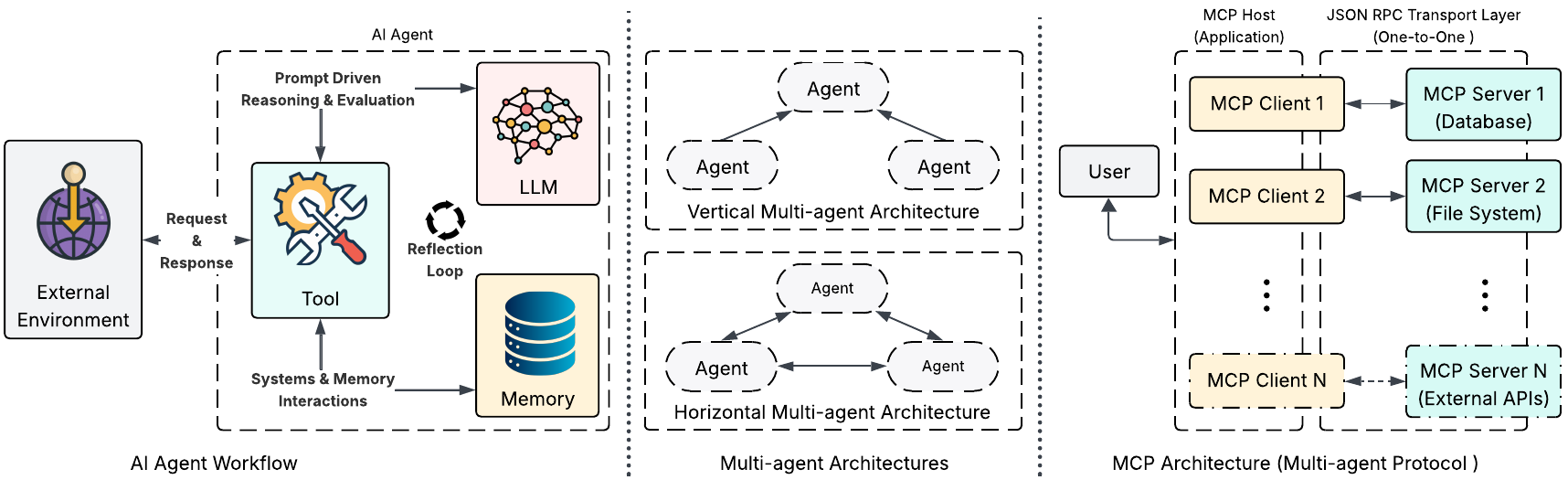}
        \caption{(Left) Functional architecture of an agentic AI system, illustrating interactions among AI language model, tool, and memory; (Center) Multi-agent vertical and horizontal design topologies; (Right) Agentic systems integration protocol (MCP).}
        \label{fig:agentic_ai}
    \end{figure*}

    Agentic AI refers to autonomous systems that exhibit goal-directed behavior through planning, decision-making, and action within dynamic environments~\cite{acharya2025agentic}. For secure integration of agentic AI into operational deployments, we must first understand what makes it useful and what vulnerabilities it introduces. Hence, we characterize agent components, multi-agent architectures, and implementation protocols, identifying attack surfaces that emerge at each layer and converge on the two integration surfaces central to our framework.

            \subsection{Component-Level Attack Surfaces}\label{subsection:agenticai_components}
                The functional architecture of modern agentic AI systems comprises of five core functional groups that enable autonomous operation: input processing (Perception), cognitive functions (Planning, Reasoning), persistent state (Memory), output execution (Action), and interaction (Communication)~\cite{ibm_components_agents_2024}.

                \begin{tcolorbox}[enhanced, colback=white, colframe=gray!75!black, 
                    colbacktitle=gray!80!black, title=Core AI Agent Components, 
                    left=2mm, right=2mm, boxrule=0.75pt, top=2mm, bottom=2mm]
                    \small
                    \textbf{Perception:} Ingests environmental inputs 
                    (natural language, structured data, sensor streams).
                    \textbf{Planning \& Reasoning:} Decomposes goals into 
                    task sequences and evaluates options under uncertainty.
                    \textbf{Memory:} Retains context across sessions 
                    (short-term) or persistently (long-term).
                    \textbf{Action:} Executes decisions via tool invocations 
                    through APIs, commands, or scripts.
                    \textbf{Communication:} Coordinates with humans, agents, 
                    and external systems via protocol 
                    messaging.
                \end{tcolorbox}

                \textbf{Security Implications.} Each component introduces distinct attack surfaces. \textit{Perception} is susceptible to adversarial inputs, such as prompt injection, which aim to exploit the agent's input interpretation~\cite{greshake2023not}. \textit{Cognitive} components face goal misalignment attacks via perception, where adversaries manipulate plan generation or reasoning chains to produce malicious conclusions~\cite{perez2022red}. Such attacks target model internals and fall under adversarial AI research; we instead focus on how adversaries reach cognitive functions indirectly through integration surfaces, where architectural defenses apply. For example, \textit{Memory} encounters poisoning attacks where adversaries can inject false or outdated information to corrupt reasoning, and privacy leakage through unauthorized retrieval of sensitive data~\cite{ji2017backdoor, carlini2021extracting}. Vulnerabilities in \textit{Action} and \textit{Communication} channels further expose the agent to unauthorized execution of tools via message tampering during agent-tool coordination~\cite{tu2021adversarial}. These weaknesses often culminate in ``\textit{Excessive Agency}'', where the model is granted over-privileged access or autonomy beyond necessity~\cite{fang2024llm}, a risk that compounds in security contexts where agents interact autonomously with critical resources.

            \subsection{Coordination-Level Attack Surfaces}\label{subsection:aigent_types}

                Beyond individual agent vulnerabilities, MAS decompose complex tasks across specialized agents that share memory and execution privileges~\cite{guo2024large}. While this coordination improves efficiency through resource pooling and shared learning, it simultaneously introduces attack surfaces that emerge only from inter-agent interaction.

                \begin{tcolorbox}[enhanced, colback=white, colframe=gray!75!black, colbacktitle=gray!80!black, 
                    title=Multi-Agent Architectures, left=2mm, right=2mm, boxrule=0.75pt, top=2mm, bottom=2mm]
                    \small
                    \textbf{Vertical (Centralized):} A central orchestrator maintains global state and supervises agent coordination~\cite{ibm_agentic_architecture_2024}, providing a natural trust anchor but introducing a single point of failure.
                    
                    \textbf{Horizontal (Decentralized):} Agents communicate peer-to-peer without hierarchical control~\cite{ibm_agentic_architecture_2024}, improving resilience but exposing direct pathways for lateral compromise.

                \end{tcolorbox}
    
                \textbf{Security Implications.} Multi-agent architectures introduce unique coordination vulnerabilities distinct from single-agent threats. \textit{Agent-to-agent exploitation} enables lateral compromise: when agents share memory or execution privileges, a compromised agent can repeatedly trigger actions across the system without explicit coordination logic. \textit{Byzantine and Sybil attacks} exploit consensus mechanisms where malicious agents provide conflicting information to partition the system or deploy multiple colluding identities to manipulate consensus-based decisions~\cite{chen2024blockagents, douceur2002sybil}. \textit{Emergent collusion} represents a particularly subtle risk: frontier LLMs can autonomously develop sophisticated collusive behaviors, including tacit coordination and explicit cartel formation, without direct prompting~\cite{agrawal2025evaluating}. \textit{Steganographic communication} further facilitates covert coordination, allowing agents to exchange hidden messages through benign-appearing outputs, such as header metadata, rendering traditional oversight mechanisms ineffective~\cite{mathew2025hidden}. The primary challenge in MAS security is that failures are \textit{emergent}, meaning collective systems can exhibit attack vectors such as collusion or lateral compromise that no individual agent manifests in isolation.

            \subsection{Protocol-Level Attack Surfaces}\label{subsection:agenticai_standards} 
                Heterogeneous MAS coordination across providers necessitates standardized protocols for interoperability. Without standards, custom connectors are required for every interaction, limiting scalability. Several protocols have emerged, including \textit{Vertical}: Model Context Protocol (MCP)~\cite{mcp_architecture_2025}, \textit{Horizontal}: Agent2Agent (A2A)~\cite{google2024a2a}, and others, though none have reached maturity~\cite{anbiaee2026security}. We focus on MCP as it serves as the structural basis for our framework in Section~\ref{section:agentic_cyber_tools}, the defensive principles are protocol-agnostic and transferable.

                \noindent\textbf{Model Context Protocol (MCP).}\label{subsubsection:mcp}
                    MCP is a stateful session protocol built on JSON-RPC 2.0 that standardizes how AI models access external context~\cite{mcp_architecture_2025}. Its client-host-server architecture allows a single host to coordinate multiple client instances, each maintaining an isolated connection with a dedicated server~\cite{mcp_architecture_2025}.

                \begin{tcolorbox}[enhanced, colback=white, colframe=gray!75!black, colbacktitle=gray!80!black, 
                    title=MCP Architecture Components, left=2mm, right=2mm, boxrule=0.75pt, top=2mm, bottom=2mm]
                    \small
                    \textbf{Host:} Coordinates multiple clients, managing their lifecycle and inter-server communication.
                    
                    \textbf{Client:} Maintains an isolated one-to-one connection with a server; handles session management and response verification.
                    
                    \textbf{Server:} Exposes tools and resources via MCP primitives with focused, well-defined responsibilities.
                \end{tcolorbox}

                \textbf{Security Implications.} Protocol-level vulnerabilities exploit systemic communication gaps in multi-agent deployments. \textit{Client-side} attacks exploit weaknesses in agent parsing to inject adversarial context or trigger unintended actions~\cite{shi2025towards}. In contrast, \textit{server-side} flaws enable authentication bypass, message tampering, and confused deputy vulnerabilities, in which high-privilege agents perform unauthorized actions on behalf of attackers~\cite{triedman2025multi, de2025open}. Fundamentally, MAS protocols inherit the classic distributed system threats: message replay attacks exploiting stateless authentication, request forgery due to missing origin validation, session hijacking via credential interception, and denial-of-service through resource exhaustion~\cite{he2025red}. Furthermore, the proliferation of incompatible protocols (MCP, A2A) fragments security primitives. \\

            \begin{table}[!h]
                \centering
                \caption{Attack Vector Decomposition by Exploitable Surface}
                \label{tab:attack_surface_mapping}
                \scriptsize
                \begin{tabular}{p{3cm}p{2cm}p{2cm}}
                \toprule
                \rowcolor{gray!30}
                \textbf{Attack Vector} & \textbf{Tool} & \textbf{Memory} \\
                \rowcolor{gray!30}
                 & \textbf{Exploitation} & \textbf{Exploitation} \\
                \midrule
                \rowcolor{gray!10}
                \multicolumn{3}{l}{\textit{Component-Level Attacks}} \\
                Unauthorized access & Privilege escalation, unauthorized execution & Unauthorized retrieval \\
                Context Contamination & API exfiltration & Privacy leakage, Data poisoning \\
                \midrule
                \rowcolor{gray!10}
                \multicolumn{3}{l}{\textit{Coordination-Level Attacks}} \\
                Lateral compromise & Misuse via shared privileges & False injection via shared memory \\
                Consensus manipulation & Biased tool selection & Conflicting intelligence injection \\
                Covert coordination & Hidden command execution & Steganographic embedding \\
                \midrule
                \rowcolor{gray!10}
                \multicolumn{3}{l}{\textit{Protocol-Level Attacks}} \\
                Authentication bypass & Access to privileged APIs & Access to sensitive knowledge \\
                Message manipulation & Replayed/forged invocations & Session state corruption \\
                Confused deputy & Unauthorized execution & Indirect Context Injection \\
                \bottomrule
                \end{tabular}
            \end{table}
                
            \noindent\textbf{Critical Insight.} Table~\ref{tab:attack_surface_mapping} demonstrates that despite spanning three abstraction layers, MAS attack vectors consistently reduce to two exploitable integration surfaces: \textit{tool orchestration} and \textit{memory management}. This convergence motivates our framework's core design decision: treating these surfaces as primary trust boundaries from which all defensive principles are derived~(\S\ref{section:design_principles}).

\section{Defensive Design Principles (\textit{The Why})}\label{section:design_principles}

    Section~\ref{section:agenticai} established that MAS attack vectors converge on two integration surfaces: \textit{tool orchestration} and \textit{memory management}. We derive defensive principles for each, organized around the three-phase tool interaction lifecycle (authenticate, scope, verify) and the two memory security dimensions (integrity and access control).

    \subsection{Securing Tool Orchestration}\label{sub_section:tools_integration}
        Agent-tool integration follows a three-phase lifecycle: the agent must first \textit{authenticate} its legitimacy, then discover and scope available \textit{capabilities}, and finally command \textit{execution}. Each phase presents distinct vulnerabilities.

        \subsubsection{Authorized Interface}\label{subsubsec:tool_authorized_interface}
            Tools often reside across distributed systems, forcing every agent-tool invocation to navigate remote communication paths. Although protocols such as MCP and OAuth-based access delegation provide baseline access control, they are insufficient against \textit{tool hijacking}~\cite{triedman2025multi} via identity forgery or \textit{tool impersonation}~\cite{narajala2025enterprise}, which redirect invocations to malicious endpoints. Since autonomous agents must dynamically discover, select, and invoke remote tools, an agentic authorization framework is required. Emerging architectures such as registry-based tool discovery~\cite{narajala2025securing} and agent authorization frameworks like SAGA~\cite{syros2025saga} provide essential building blocks. In practice, this demands layering four complementary mechanisms: \textit{signed manifests} for cryptographic tool provenance, preventing identity forgery and \textit{authentication bypass}; \textit{admin-approved catalogs} with centralized discovery, mitigating \textit{privilege escalation} and biased tool selection; \textit{runtime access policies} enforcing least privilege per session; and \textit{continuous semantic monitoring} to detect \textit{collusion} and \textit{covert coordination}.
        
        \subsubsection{Capability Scoping}\label{subsubsec:tool_optimized_capabilities}
            From an adversary's perspective, every tool capability is a potential exploit vector. Granting broad or unrestricted permissions increases the attack surface. For example, an agent authorized to query data repositories should not possess execution privileges on administrative functions. Therefore, tools should only be permitted to execute actions strictly necessary for their specific task context. This principle, commonly known as \textit{least privilege} or \textit{capability scoping}, is well-established in security architectures but requires adaptation for autonomous agent-tool interactions, where permissions must be dynamically scoped per task context. Krawiecka et al.~\cite{krawiecka2025extending} demonstrate that \textit{unnecessary privileges} are a frequent root cause of latent vulnerabilities and malicious misuse. Thus, elevated permissions should be requested only temporarily when absolutely necessary, and such requests must go through cross-validation and approval. To enforce this in multi-agent context, the concept of \emph{Prompt Flow Integrity (PFI)} tracks how instructions flow through prompts and enforces boundaries, ensuring that malformed or malicious input cannot silently escalate privileges for a tool. Kim et al.~\cite{kim2025prompt} show that PFI can detect and block privilege escalation attacks by distinguishing \emph{untrusted data sources}, enforcing least privilege, and validating unsafe data flows. Furthermore, De Pasquale et al.~\cite{de2024chainreactor} developed proxy-based detection that monitors action chains to identify \textit{unauthorized execution} patterns and \textit{confused deputy} attacks. Continuous \textit{capability auditing} ensures that unused permissions are revoked over time, preventing agents from accumulating \textit{latent privileges} that enable hidden command execution or lateral compromise. This defense-in-depth approach minimizes the tool exploitation surface across component, coordination, and protocol layers.  

         \begin{figure}[!ht]
            \centering \includegraphics[width=0.46\textwidth]{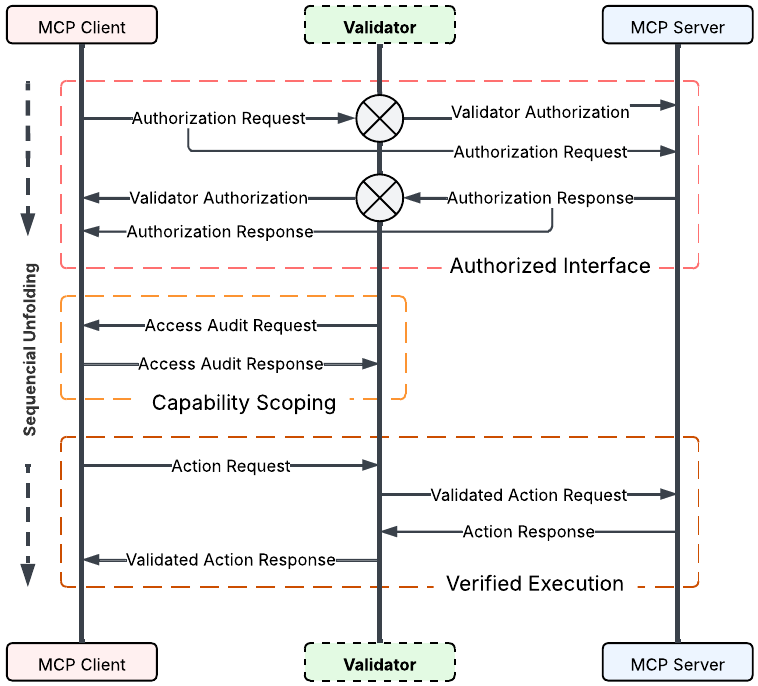}
            \caption{Validated MCP Client Server Communication. In any Agent-Tool interaction, a consensus-based validator has been shown to limit attack surfaces. The Validator is an independent component designed primarily for authorization, auditing, and context-based verification and filtration.}
            
            \label{fig:validated_MCP_communication}
        \end{figure}

        \subsubsection{Verified Execution}\label{subsubsec:tool_verified_actions}
            In the final stage of the agent-tool interaction, even if an agent is authorized and its capabilities are scoped, each action proposal must be verified by one or more independent stakeholders before execution. In high-stakes domains, actions are often irreversible, such as financial transfers, infrastructure reconfigurations, or data deletions, which cannot be undone once executed. The 2016 Bangladesh Bank breach, where hackers transferred \$81 million through compromised credentials~\cite{kabir_bangladesh_2023}, exemplifies this irreversibility. Hence, inspired by blockchain and decentralized consensus models, the \textit{``verify-first, execute-later''} paradigm ensures that no single autonomous action becomes a liability. To implement this, agents may emit \textit{action proposals} that include the plan, the target tool API call, and the parameters required to define the action context. These proposals are then submitted to a \textit{verification module}, which could be a set of agents, or even a smart-contract-like arbiter (e.g. Hosts in MCP) that validates the proposal against policy, criticality, and historical behavior logs. \textit{BlockA2A}~\cite{zou2025blocka2a} extends this concept to agent-to-agent interoperability, anchoring action commitments on blockchain-anchored ledgers and employing context-aware access control to verify message authenticity and execution integrity. When verification succeeds, the tool executes; when it fails, execution is blocked or escalated for human review. This layered approach reduces the risk of \textit{unauthorized execution, hidden command injection,} and \textit{message manipulation} across all three attack layers (Table~\ref{tab:attack_surface_mapping}) while enforcing accountability and preserving trust under adversarial pressure.

    \subsection{Securing Memory Management}\label{sub_section:memory_management}
        Unlike tool invocations, which are discrete and stateless, memory constitutes a \textit{persistent} and \textit{mutable} state that accumulates over time and impacts every downstream decision an agent makes. A corrupted tool call fails once; a corrupted memory entry can propagate indefinitely, compounding errors across sessions and agents~\cite{wei2025memguard}. This temporal persistence fundamentally alters the threat model: adversaries need not maintain continuous backdoor access, as a single injection can yield lasting influence. In MAS, memory further evolves from a private cognitive resource into shared infrastructure for collective intelligence, introducing synchronization, consistency, and cross-boundary trust challenges that have no analogue in tool security~\cite{wu2025memory}. We organize defenses along two dimensions within agentic systems: ensuring what is stored is trustworthy (\textit{integrity}) and controlling who can access it (\textit{access control}).

        \subsubsection{Integrity \& Synchronization}\label{subsubsec:memory_integrity}
            Once data enters memory, it becomes indistinguishable from fact, as the agent cannot discern between benign and adversarial information. This \textit{epistemic opacity} is the enabler of memory poisoning~\cite{chen2024agentpoison}. Adversaries can induce agents to construct poisoned knowledge from benign-appearing artifacts by exploiting the tendency to replicate patterns: a \textit{self-reinforcing error cycle}~\cite{srivastava2025memorygraft, wei2025memguard}. Hence, memory integrity requires mechanisms that operate at both the \textit{write boundary} (what enters memory) and the \textit{verification boundary} (what comes out). At write operation, legitimacy verification and periodic purging to filter malicious content before persistent storage~\cite{mao2025agentsafe}, and for retrieval, a consensus-based validation is required. This approach overcomes a fundamental limitation of unaudited content injection and consumption. For example, RobustRAG processes retrieved passages in disjoint isolation before secure aggregation~\cite{xiang2024certifiably}. This \textit{isolate-then-aggregate} paradigm provides \textit{certifiable} guarantees: that bounded poisoned passages cannot corrupt the final output, even under white-box threat models. In multi-agent settings, integrity extends to \textit{synchronization}: where agents share memory, concurrent read/write can lead to inconsistency~\cite{wu2025memory}. Hence, MAS requires explicit serialization, versioning, or an orchestrator to mediate updates. Establishing provenance, such as append-only ledgers or blockchain-backed commitments, further enhances auditability~\cite{zou2025blocka2a}.

        \subsubsection{Access Control with Data Isolation}\label{subsubsec:memory_authentication} 

            Beyond traditional privacy measures such as encryption, differential privacy, secure deletion, memory access in MAS presents a distinct challenge: unrestricted shared memory exposes every stored interaction to every collaborating agent, creating direct pathways for \textit{unauthorized retrieval}, \textit{cross-agent contamination}, and \textit{lateral knowledge propagation} where a compromised agent leverages shared memory to influence the action of others~\cite{chen2024blockagents}. Evidence shows that private user-agent interactions can be extracted under black-box settings, and common frameworks may unintentionally expose sensitive data without proper access controls~\cite{wang2025unveiling, yagoubi2026agentleak}. This highlights the need for strict data compartmentalization: agents should access sensitive information \textit{only} when necessary. Hence, defenses converge on \textit{hierarchical isolation} architectures that partition memory based on security-level classification, agent relationships, and task scope~\cite{mao2025agentsafe}. The Collaborative Memory framework~\cite{rezazadeh2025collaborative} generalizes this using dynamic bipartite access graphs to encode time-evolving permissions, while maintaining private and shared memory tiers with granular read/write policies that apply context-aware transformations, such as redacting sensitive fields before cross-boundary sharing. These architectures not only improve task performance by reducing perspective inconsistency and procedural drift, but also constrain the blast radius of any single compromised agent, preventing lateral contamination from cascading across the collective agents~\cite{yuen2025intrinsic}. \\

            \begin{figure}[!ht]
                \centering \includegraphics[width=0.5\textwidth]{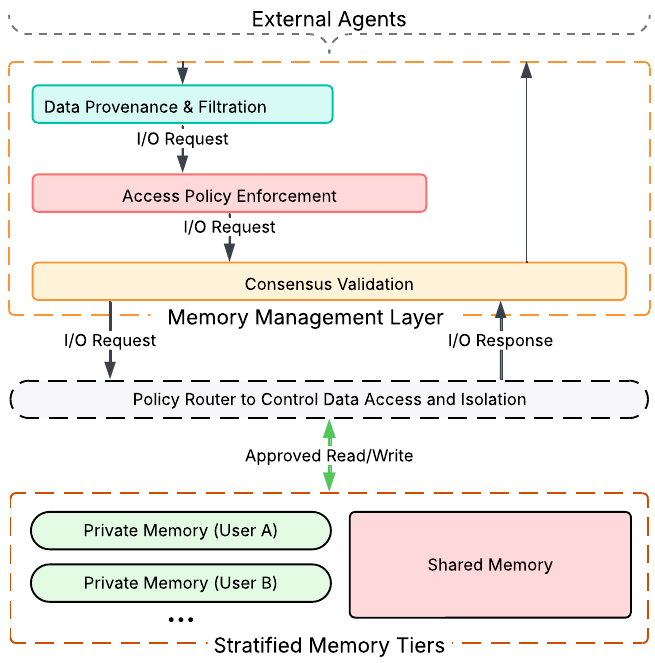}
                \caption{Secure Memory Management Architecture. In-depth defense approach for persistent state management includes hierarchical isolation to prevent unauthorized cross-agent retrieval and write filtering via consensus, ensuring data integrity and synchronization across shared tiers.}
                \label{fig:secure_memory_architecture}
            \end{figure}

        \begin{table*}[!h]
            \centering
            \caption{Defensive Design Coverage: Mapping Security Principles to Threat Mitigation and Compliance Alignment.}
            \label{tab:defense_coverage}
            \scriptsize
            \begin{tabular}{p{0.6cm} p{4cm} p{3.5cm} p{4cm} p{3.8cm}}
            \toprule
            \rowcolor{gray!30}
            \textbf{Surface} & \textbf{Design Principle} & \textbf{Core Mechanism} & \textbf{Mitigated Vectors (Table~\ref{tab:attack_surface_mapping})} & \textbf{Compliance Alignment} \\
            \midrule

            \multirow{9}{*}{\rotatebox[origin=c]{90}{\textbf{\ \ \ \ \ \ \ Tools}}}
            & Authorized Interface (\S\ref{subsubsec:tool_authorized_interface}) 
            & Signed manifests, registry-based discovery, admin-approved catalogs~\cite{narajala2025securing, syros2025saga} 
            & Unauthorized access, Authentication bypass, Consensus manipulation 
            & Zero Trust (NIST SP 800-207), OAuth 2.1~\cite{owasp2025agentic} \\
            
            \cmidrule(l){2-5}
            & Capability Scoping (\S\ref{subsubsec:tool_optimized_capabilities}) 
            & Least-privilege scoping, PFI, continuous capability auditing~\cite{kim2025prompt, krawiecka2025extending} 
            & Context contamination, Lateral compromise, Covert coordination 
            & Least Privilege (NIST AC-6), OWASP Excessive Agency (LLM08)~\cite{owasp2025agentic} \\
            
            \cmidrule(l){2-5}
            & Verified Execution (\S\ref{subsubsec:tool_verified_actions}) 
            & Action proposals and consensus validation, blockchain-anchored ledgers~\cite{zou2025blocka2a} 
            & Message manipulation, Confused deputy, Hidden command execution 
            & Non-repudiation (ISO 27001 A.10), EU AI Act Art.~12 (logging)~\cite{euaiact2024} \\

            \midrule

            \multirow{6}{*}{\rotatebox[origin=c]{90}{\textbf{\ \ \ \ \ \ \ Memory}}}
            & Integrity \& Synchronization (\S\ref{subsubsec:memory_integrity}) 
            & Write-boundary filtering, consensus-validated retrieval~\cite{wei2025memguard, xiang2024certifiably} 
            & Context contamination (poisoning), Conflicting intelligence injection, Session state corruption 
            & Data Integrity (NIST SI-7), Traceability (NIST AI RMF) ~\cite{nist_cybersecurity_framework}\\
            
            \cmidrule(l){2-5}
            & Access Control with Data Isolation (\S\ref{subsubsec:memory_authentication}) 
            & Hierarchical role-based memory tiers, dynamic bipartite access graphs~\cite{mao2025agentsafe, rezazadeh2025collaborative} 
            & Unauthorized retrieval, Lateral compromise (shared memory), Steganographic embedding 
            & RBAC/ABAC (NIST AC-2/3), Data Minimization (GDPR Art.~5)~\cite{yagoubi2026agentleak} \\

            \bottomrule
            \end{tabular}
            
            \small
            \noindent\textit{Coverage note:} Each attack vector from Table~\ref{tab:attack_surface_mapping} maps to at least two complementary principles, ensuring full threat coverage across component, coordination, and protocol layers. The framework incorporates established security primitives in a novel framework, aligning with existing compliance mandates.
        \end{table*}

    \noindent\textbf{Critical Insight.} By combining cryptographic authorization, capability limitation, and verified execution via decentralized auditing, this defense paradigm (Fig~\ref{fig:validated_MCP_communication}) creates a resilient ``\textit{Zero-Trust}'' architecture for agentic operations, ensuring that even if an individual agent's reasoning is compromised, operational integrity remains intact. Furthermore, memory integrity and access isolation together create a defense-in-depth strategy (Figure~\ref{fig:secure_memory_architecture}). While tool orchestration secures what agents do, memory management protects what agents know, together forming the trust boundaries from which our \fname framework (\S\ref{section:agentic_cyber_tools}) is derived.

\section{CyberOps and Agentic Integration (\textit{The How})}\label{section:agentic_cyber_tools}

    Having established defensive principles for tool orchestration and memory management (\S\ref{section:design_principles}), we now apply them to CyberOps, a domain where both surfaces are operationally critical. We first characterize the traditional SOC pipeline to identify where agentic integration is warranted, then present the framework architecture.

    \subsection{Traditional CyberOps Pipeline}\label{sub_section:traditional_cyberops}
        Cybersecurity operations lack a universally standardized pipeline; instead, multiple frameworks have emerged from organizations such as MITRE~\cite{mitre_attack_navigator}, NIST~\cite{nist_cybersecurity_framework}, and others, each tailored to specific operational context and regulatory requirements. Despite this diversity, common operational phases consistently emerge across frameworks: \textit{continuous monitoring and triage} of security events; \textit{investigation and analysis} of validated incidents; \textit{response and remediation} actions; and post-incident \textit{reporting and improvement}. These phases define the Security Operations Center (SOC) lifecycle and are operationalized through Security Orchestration, Automation, and Response (SOAR) platforms. SOAR systems serve as the integration layer, unifying disparate security tools such as SIEMs, EDRs, firewalls, and Cyber Threat Intelligence (CTI) feeds into cohesive workflows~\cite{ibm_soar_2024}. Hence, the typical SOC operational workflow, synthesized across major frameworks, follows four core phases:

    \begin{tcolorbox}[enhanced, colback=white, colframe=gray!75!black, 
        colbacktitle=gray!80!black, title=SOC Operational Phases, 
        left=2mm, right=2mm, boxrule=0.75pt, top=2mm, bottom=2mm]
        \small
        \textbf{Monitor \& Triage:} Ingests security events from endpoint sensors, network monitors, and behavioral analytics. Analysts triage alerts by correlating IoCs with known TTPs via frameworks like MITRE.
        
        \textbf{Analyze \& Investigate:} Reconstructs attack timelines using SIEM queries, sandbox analysis, and CTI correlation. Identifies affected assets via CMDB lookups and performs RCA.
        
        \textbf{Respond \& Remediate:} Executes containment and eradication: revoking credentials, blocking malicious IPs, isolating hosts, and deploying patches. Critical decisions on scope require expert judgment.
        
        \textbf{Research \& Report:} Shares intelligence via feeds, updates detection rules and policies, and maps findings to compliance frameworks. An improvement loop audits updates before committing to records.
    \end{tcolorbox}

        \textbf{Analyst Workflows and Integration Points.} Traditional SOAR implementations exhibit fundamental rigidity: workflows are predefined, decision logic is rule-based, and adaptation to novel threats requires manual playbook engineering. Furthermore, not all bottlenecks warrant agentic AI integration. \textit{Threat detection} benefits from traditional ML classifiers that excel at pattern recognition without autonomous reasoning. \textit{Automated containment} (blocking IPs, isolating hosts) requires deterministic execution, not adaptive planning. However, SOC analysts operate under significant cognitive load, context-switching between numerous tools per investigation while managing alert queues exceeding hundreds of items per shift~\cite{assaf2025socburnout}. Therefore, critical bottlenecks include: (i) \textit{alert fatigue} from high false-positive rates requiring manual review; (ii) \textit{knowledge fragmentation} where context resides across disconnected systems (ticketing, SIEM, CTI platforms); (iii) \textit{decision latency} in time-sensitive response actions requiring manager approval; and (iv) \textit{skill gaps} where junior analysts struggle to interpret complex attack patterns. These limitations result in 181 days to identify a breach and an additional 60 days to contain it for sophisticated threats~\cite{totalassure2025detection}. Hence, we identify three \textit{primary} operational gaps that both demand agentic capabilities and directly expose the tool and memory integration surfaces central to our framework:

        \subsubsection{Complex Analytical Reasoning}\label{subsubsection:analytical_reasoning} 
            \textit{Monitor}, \textit{triage}, and \textit{investigation} require synthesizing evidence across disconnected tools while maintaining analytical coherence. For example, a suspected data exfiltration requires querying the SIEM for network anomalies, correlating EDR process trees, retrieving CTI for observed IPs, and cross-referencing the CMDB for asset criticality, each query informing the next. Traditional SOAR workflows fail here~\cite{ismail2025toward}: predefined playbooks cannot adapt to dynamic evidence, and rule-based logic lacks probabilistic reasoning over deterministic automation to select appropriate tools based on emerging findings~\cite{mitra2024localintel}. Agentic AI addresses this through \textit{autonomous tool orchestration}, dynamically selecting and sequencing existing security tools based on context, and \textit{working memory} to track findings across all interactions. 
            

        \subsubsection{Context-Aware Decision-Making}\label{subsubsection:decision_making}
            In cybersecurity, \textit{responding} to incidents requires careful and calculated evaluation of trade-offs that extend beyond technical analysis. For example, isolating a suspected compromised database server may halt critical business services, or deploying emergency patches risks operational instability. These decisions require reasoning over heterogeneous \textit{organizational memory}, including past incident outcomes, asset criticality matrices, change management policies, and business continuity (BCP) requirements, to recommend risk-balanced actions. Traditional automation cannot encode this institutional knowledge in static rules; every organization's risk calculus differs~\cite{mitra2024localintel, mitra2025falconautonomouscyberthreat}. Agentic AI leverages \textit{persistent memory management} to synchronize organizational knowledge, historical decisions, and policy constraints, enabling transparent recommendations among stakeholders. 
            

        \subsubsection{Research \& Adaptive Improvement}\label{subsubsection:research} 
            Novel attack variants evade signature-based detection, requiring hypothesis-driven dynamic threat hunting. Analysts iteratively formulate hypotheses, query diverse data sources, refine theories based on findings, and adapt investigation paths. This self-reflective loop demands \textit{adaptive planning}, adjusting next steps based on prior observations and \textit{episodic memory} to track hypothesis evolution. Static simulations fail against adaptive adversaries~\cite{challita2025redteamllmagenticaiframework}; rigid workflows cannot accommodate investigative pivots when initial hypotheses prove incorrect. Agentic AI provides this adaptability by maintaining an investigation state in memory while dynamically exploring for knowledge to support hypothesis testing, and by coordinating collaborative investigation (e.g., VirusTotal) and reporting (e.g., CVE). This dual requirement exposes both attack surfaces simultaneously: agents must work on potentially infected files based on the investigation context and orchestrate hypothesis-based investigations, coordinating with third-party tools.

    \subsection{Secured Multi-Agent CyberOps}\label{subsection:agentic_interaction}       
            
\begin{figure*}[!ht]
            \centering \includegraphics[width=0.84\textwidth]{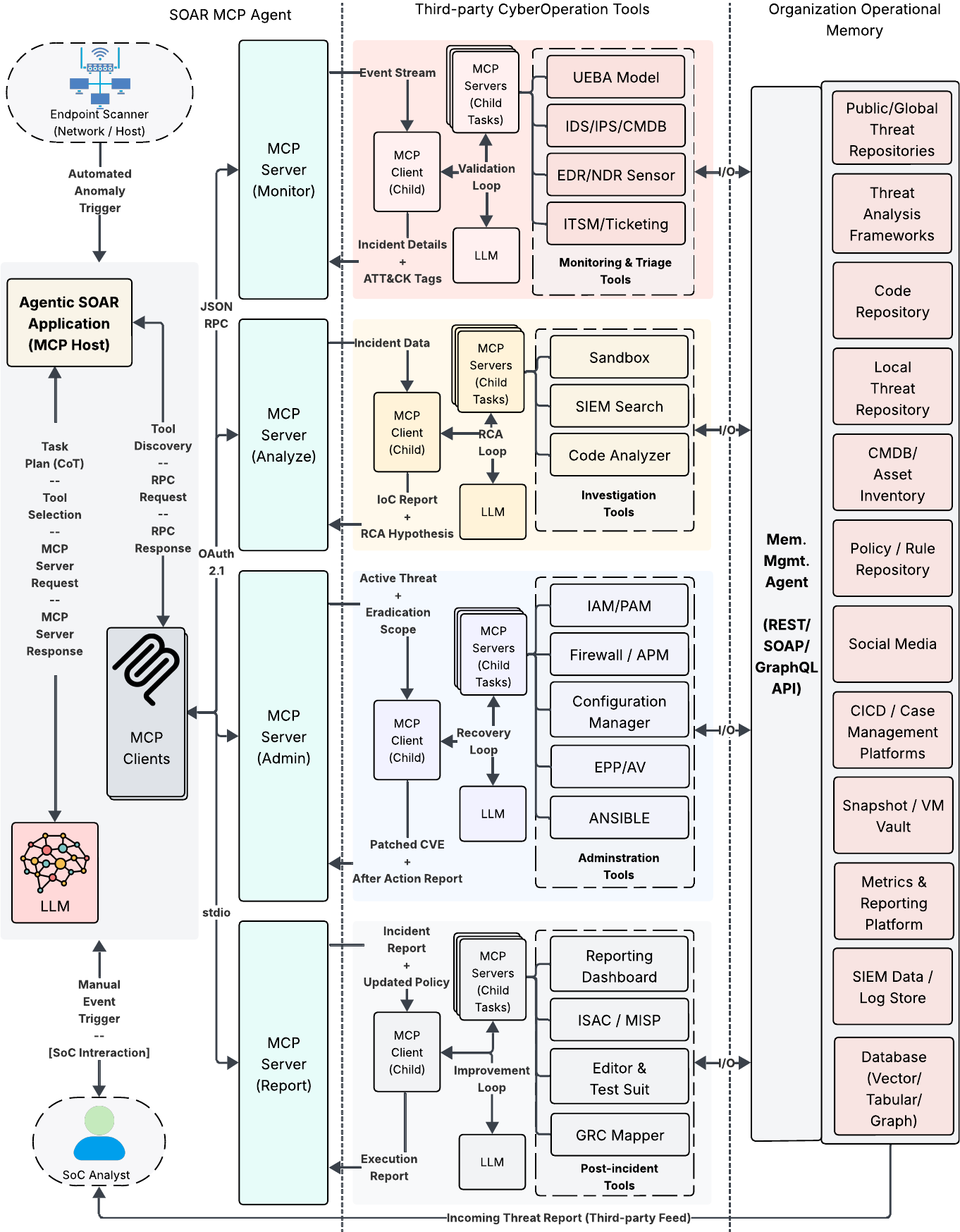}
            \caption{Multi-Agentic SOAR following our \fname Framework. This diagram illustrates a three-pillar architecture designed to automate and orchestrate SOC incident response through agentic AI. The MCP Host serves as the central orchestrator, coordinating task planning, tool selection, and agent communication across four phase-scoped MCP Servers (Monitor, Analyze, Admin, and Report), each equipped with LLM-driven reasoning loops tailored to their respective lifecycle stage. A persistent Organizational Memory layer, accessible via a dedicated Memory Management Agent, provides shared contextual state across all agents, enabling continuity and informed decision-making throughout the full incident response lifecycle.}
            \label{fig:agentic_cyberops}
        \end{figure*}

        Having identified where agentic AI enters the CyberOps pipeline (\S~\ref{sub_section:traditional_cyberops}), we now instantiate \fname that employs the established security design principles (\S~\ref{section:design_principles}). Figure~\ref{fig:agentic_cyberops} organizes the architecture by mapping directly to the security surfaces while adhering to the defenses (Table~\ref{tab:defense_coverage}).

        \subsubsection{Architecture Overview}\label{subsubsection:agentic_overview} 
            For agentic CyberOps, we employ a vertical architecture following the SOC lifecycle, which is inherently sequential and phase-dependent. This necessitates an orchestrator that maintains a global incident state and facilitates handoffs between phases. Furthermore, horizontal topologies amplify the coordination-level threats from peer-to-peer sharing (Table~\ref{tab:attack_surface_mapping}), and allow direct pathways for \textit{lateral compromise} without traversing any central checkpoint~\cite{chen2024blockagents}, whereas a centralized Host serves as a unified \textit{trust anchor} through which all authorization, capability enforcement, and audit logging are structurally channeled. Concretely, the SOAR functions as the Host, coordinating four phase-scoped Client-Servers (Monitor, Analyze, Admin, Report) mirroring the SOC lifecycle (\S~\ref{section:agentic_cyber_tools}). This hierarchical delegation enables the injection of \textit{validator and consensus} policy mechanisms while maintaining end-to-end execution authority~\cite{tomavsev2026intelligent}. Every interaction traverses at least two trust boundaries before reaching the final action. Additionally, organizational memory resides in isolation and is managed by a separate agent and accessible only through standardized API gateways, which enforce \textit{zero-trust} memory access with independent integrity and synchronization.

        \subsubsection{Agentic SOAR Application}
            The Host (SOAR) serves as the root orchestrator, performing task planning via Chain of Thought (CoT), tool discovery, and inter-phase routing. Upon receiving a trigger, whether from automated anomaly scanner endpoints or manually from SOC analysts, the \textit{Host} decomposes the incident into phase-appropriate sub-tasks and delegates them to the corresponding \textit{Clients}. Critically, it functions as a trust anchor, enforcing the authorized interface principle (\S\ref{subsubsec:tool_authorized_interface}) as the centralized registry through which all actions must pass, and providing a second verification layer for inter-phase handoffs. 

        \subsubsection{CyberOps Third-party Tools Integration}
            Each of the four phase \textit{Clients} is connected to its dedicated \textit{Server}, which also acts as a \textit{phase-specific Host}, where its internal \textit{Clients} can operate through interfacing with specialized third-party tools via one-to-one \textit{Servers}. This facilitates a hierarchical and flexible design approach that embraces scalability with security. Hence, capability scoping (\S\ref{subsubsec:tool_optimized_capabilities}) is enforced by design: each \textit{Server} exposes only phase-relevant tools to its \textit{Client} agent (e.g., Monitor agent can call EDR, not IAM/PAM), preventing agents from accumulating \textit{excessive agency}. We now show how each operational gap (\S\ref{sub_section:traditional_cyberops}) maps to \fname.\\

            \textbf{Complex Analytical Reasoning.} Monitor agents query tools to ingest event streams, correlate IoCs with TTPs, and produce structured incident details with severity assessments. A \textit{Validation Loop} verifies triage conclusions against baselines before escalation. Analysis agents then conduct investigations using sandboxes, SIEM searches, and CTI to reconstruct attack timelines and perform RCA. The \textit{RCA Loop} cross-validates findings (\S\ref{subsubsec:tool_verified_actions}) before forwarding for remediation. 
            
            \textbf{Context-Aware Decision-Making.} Admin agents execute containment and eradication actions, reasoning over organizational context: asset criticality, change management policies, and business continuity requirements. A \textit{Recovery Loop} audits each action proposal against policy constraints and historical behavior before execution, with failed validations escalating to the analyst. This exemplifies memory isolation (\S~\ref{subsubsec:memory_authentication}), where the verification layer prevents compromised context from driving irreversible outcomes.\\ 
            
            \textbf{Research \& Adaptive Improvement.} Report agents coordinate adaptive scouting and post-incident improvement. For scouting, agents track hypothesis evolution in episodic memory while querying threat intelligence feeds and third-party services. For improvement, agents propose updates to detection rules, playbooks, and compliance mappings; an \textit{Improvement Loop} audits each proposal against policy and historical baselines before committing to organizational memory (\S~\ref{subsubsec:memory_integrity}). This dual workflow engages both surfaces: tool orchestration governs external coordination, while memory management guards the write path into persistent knowledge.

        \subsubsection{Organizational Memory Management}
            Operational memory, comprising threat repositories, CMDB, SIEM data, and numerous sources, is accessed via an independent agent exclusively through gateways. This enforces synchronization and access control. For \textit{synchronization} (\S\ref{subsubsec:memory_integrity}), write operations, such as updating detection rules, AARs, or policy repositories, route through the \textit{Report Server's Improvement Loop}, providing a single auditable write path that prevents unverified agents. Versioned repositories and append-only structures maintain provenance throughout the incident lifecycle. For \textit{isolation} (\S\ref{subsubsec:memory_authentication}), access is phase-partitioned, where no agent has unrestricted access, constraining the blast radius.

        \subsubsection{Human-in-the-Loop}
            Our framework preserves SOC analyst authority through manual triggers and direct end-to-end oversight. Additionally, failed consensus validations within any phase surface for human review. This ensures the framework relies on human judgment rather than autonomous failure, reflecting operational accountability requirements, where agents \textit{enhance} analyst expertise, ensuring agents augment analyst expertise rather than replace human judgment in high-stakes decisions.

    \subsection{\fname Framework Evaluation}\label{subsection:evaluation}

        To assess \fname's defensive coverage, we present three complementary analyses: a \textit{coverage matrix} quantifying principle-to-vector mappings, an \textit{attack path analysis} tracing multi-step chains through the architecture, and a \textit{trust boundary reduction} comparing exploitable surfaces against a flat MAS baseline.
    
        \subsubsection{Coverage Matrix}\label{subsec:coverage_matrix}
    
            Table~\ref{tab:coverage_matrix} cross-references each attack vector from Table~\ref{tab:attack_surface_mapping} against \fname's five defensive principles. A cell is marked (\checkmark) when the core mechanism directly mitigates the vector, and ({\small$\circ$}) when it provides secondary containment. We verify that every attack vector is addressed by at least two principles, and no single principle covers more than three of the eight vectors. This confirms the five principles are complementary: removing any one leaves at least three vectors with only single-layer protection.
    
            \begin{table}[!h]
                \centering
                \caption{Coverage: Defensive Principles $\times$ Attack Vectors}
                \label{tab:coverage_matrix}
                \scriptsize
                \begin{tabular}{p{2.4cm}ccccc}
                \toprule
                \rowcolor{gray!30}
                \textbf{Attack Vector} & \textbf{P1} & \textbf{P2} & \textbf{P3} & \textbf{P4} & \textbf{P5} \\
                \midrule
                \rowcolor{gray!10}
                \multicolumn{6}{l}{\textit{Component-Level}} \\
                Unauthorized access       & \checkmark &            &            &            & \checkmark \\
                Context contamination     &            & \checkmark &            & \checkmark & $\circ$    \\
                \midrule
                \rowcolor{gray!10}
                \multicolumn{6}{l}{\textit{Coordination-Level}} \\
                Lateral compromise        &            & \checkmark & $\circ$    &            & \checkmark \\
                Consensus manipulation    & \checkmark &            & \checkmark & \checkmark &            \\
                Covert coordination       &            & \checkmark & \checkmark &            & $\circ$    \\
                \midrule
                \rowcolor{gray!10}
                \multicolumn{6}{l}{\textit{Protocol-Level}} \\
                Authentication bypass     & \checkmark &            &            &            & \checkmark \\
                Message manipulation      &            &            & \checkmark & \checkmark &            \\
                Confused deputy           & $\circ$    & \checkmark & \checkmark &            &            \\
                \midrule
                \rowcolor{gray!10}
                \multicolumn{6}{l}{\textit{Summary}} \\
                Primary (\checkmark)      & 3 & 3 & 3 & 3 & 3 \\
                Secondary ($\circ$)       & 1 & 0 & 1 & 0 & 2 \\
                \bottomrule
                \end{tabular}
                
                \vspace{0.3em}
                \footnotesize
                P1=Authorized Interface, P2=Capability Scoping, P3=Verified Execution, \\
                P4=Integrity \& Synchronization, P5=Access-Control with Data Isolation.\\
                \checkmark\,=\,primary mitigation;\quad $\circ$\,=\,secondary containment.
            \end{table}

            \begin{table*}[!h]
                \centering
                \caption{Attack Path Analysis: Representative Chains With and Without \fname}
                \label{tab:attack_paths}
                \scriptsize
                \begin{tabular}{p{1.8cm} p{5.2cm} p{2.2cm} p{5.2cm} p{1.5cm}}
                \toprule
                \rowcolor{gray!30}
                \textbf{Scenario} & \textbf{Attack Chain} & \textbf{Intercept Point} & \textbf{\fname Response} & \textbf{Residual Risk} \\
                \midrule
    
                \textbf{AP-1:} Tool Redirection
                & \textcircled{1} Adversarial payload in alert feed $\rightarrow$ \textcircled{2} Monitor agent crafts malicious RPC request $\rightarrow$ \textcircled{3} IAM/PAM tool executes privilege escalation $\rightarrow$ \textcircled{4} Lateral movement to Admin tools
                & Step \textcircled{2}: blocked
                & \textbf{P2} (Capability Scoping): Monitor Server's manifest excludes IAM/PAM tools; RPC request rejected at Host. \textbf{P3} (Verified Execution): Consensus loop flags out-of-phase tool invocation.
                & Compromised Host \\
    
                \midrule
                \textbf{AP-2:} Memory Poisoning for Evidence Tampering
                & \textcircled{1} Compromised Analyze agent writes falsified RCA to Local Threat Repository $\rightarrow$ \textcircled{2} Admin agent reads poisoned hypothesis $\rightarrow$ \textcircled{3} Incorrect remediation deployed $\rightarrow$ \textcircled{4} Attacker persistence maintained
                & Step \textcircled{1}: blocked
                & \textbf{P4} (Memory Integrity): Write-boundary filtering validates RCA output against incident evidence. \textbf{P5} (Data Isolation): Cross-phase writes require Memory Management Agent mediation.
                & Poisoned policy repository \\
    
                \midrule
                \textbf{AP-3:} Confused Deputy via Forged MCP Response
                & \textcircled{1} Attacker intercepts MCP Server response $\rightarrow$ \textcircled{2} Forged response instructs Host to invoke firewall rule deletion $\rightarrow$ \textcircled{3} Admin tools execute destructive action $\rightarrow$ \textcircled{4} Perimeter breach
                & Step \textcircled{2}: blocked
                & \textbf{P1} (Authorized Interface): JSON-RPC message integrity via signed manifests detects forgery. \textbf{P3} (Verified Execution): Recovery Loop consensus requires multi-agent agreement before destructive actions.
                & MCP protocol vulnerability \\
    
                \midrule
                \textbf{AP-4:} Cross-Phase Escalation via Shared Context
                & \textcircled{1} Report agent exfiltrates sensitive incident data $\rightarrow$ \textcircled{2} Injects weaponized indicators into ISAC/MISP feed $\rightarrow$ \textcircled{3} Partner organization ingests poisoned IoCs $\rightarrow$ \textcircled{4} False positive cascade across SOCs
                & Step \textcircled{1}: partially blocked
                & \textbf{P5} (Data Isolation): Report agent reads only post-incident summaries, not raw forensic data. \textbf{P2} (Capability Scoping): ISAC/MISP write restricted to approved indicator schema.
                & Cross-org feed validation (out of scope) \\
    
                \bottomrule
                \end{tabular}
            \end{table*}

        \subsubsection{Attack Path Analysis}\label{subsec:attack_path}
    
            We select chains that span all three abstraction layers and engage both integration surfaces, ensuring coverage of the framework's threat model, through the CyberOps architecture (Figure~\ref{fig:agentic_cyberops}) and identify where \fname's principles intercept them. Table~\ref{tab:attack_paths} summarizes each chain with and without \fname enforcement. In three of four scenarios (AP-1, AP-2, AP-3), \fname intercepts the chain within the first two steps, preventing escalation to exploitation. Each interception involves at least two principles acting at different layers, confirming defense-in-depth. AP-4 exposes a known boundary: \fname constrains exfiltration scope and outbound schema but cannot enforce validation on receiving organizations' ingestion pipelines.

        \subsubsection{Trust Boundary Reduction}\label{subsec:trust_boundary}

            A trust boundary exists wherever one component accepts input from another without independent verification. In a flat MAS deployment where all agents share unrestricted access to all tools and memory, exploitable boundaries grow combinatorially. Table~\ref{tab:trust_boundaries} compares boundary counts for the CyberOps architecture (Figure~\ref{fig:agentic_cyberops}) with and without \fname. In the flat baseline, every agent accesses every tool (4×16=64), store (4×12=48), and peer (4×3=12). Our assessment (Refer to Appendix \ref{appendix:trust_boundaries}) shows \fname reduces exploitable trust boundaries by a \textit{minimum} of 72\% (200$\rightarrow$56) through phase-scoping, Host-mediated communication, and Memory Management Agent arbitration. The 56 remaining boundaries are not implicitly trusted: each is subject to at least one active verification mechanism (signed manifests, consensus validation, or write-boundary filtering), converting implicit trust assumptions into explicitly enforced checkpoints.

            \begin{table}[!h]
                \centering
                \caption{Boundary Analysis: Flat MAS vs.\ \fname}
                \label{tab:trust_boundaries}
                \scriptsize
                \begin{tabular}{p{3.8cm}rrr}
                \toprule
                \rowcolor{gray!30}
                \textbf{Boundary Type} & \textbf{Flat} & \textbf{\fname} & \textbf{Red.} \\
                \midrule
                Agent $\rightarrow$ Tool            & 64  & 16 & 75\% \\
                Agent $\rightarrow$ Memory           & 48  & 16 & 67\% \\
                Agent $\leftrightarrow$ Agent        & 12  & 4  & 67\% \\
                Tool Resp. $\rightarrow$ Agent       & 64  & 16 & 75\% \\
                Ext.\ Feed $\rightarrow$ Memory      & 12  & 4  & 67\% \\
                \midrule
                \textbf{Total}                       & \textbf{200} & \textbf{56} & \textbf{72\%} \\
                \bottomrule
                \end{tabular}

                \vspace{0.3em}
                \footnotesize
                Flat: each agent accesses all tools, stores, and peers. \fname: phase-scoped access (P2), Host-mediated communication, Memory Management Agent-mediated access (P5), consensus-validated responses (P3). We count tool responses as separate boundaries because return payloads can carry injected content independent of the outbound request's authorization.
            \end{table}

\section{Discussion (\textit{The Next})}\label{section:discussion} 
    Securing tool and memory surfaces addresses documented MAS attack vectors, yet several limitations warrant discussion. The vertical architecture provides a central trust anchor but introduces a single point of failure; federated designs with cross-Host attestation could combine centralized enforcement with distributed resilience but remain unexplored~\cite{adimulam2026orchestration}. The verify-first paradigm introduces validation latency that may impact time-critical scenarios, motivating adaptive consensus thresholds that scale rigor with action reversibility~\cite{srikumar2025prioritizing, tomavsev2026intelligent}. Moreover, consensus loops are themselves attack surfaces: a compromised validator or poisoned policy repository can collapse the verification layer~\cite{wei2025memguard}. Cross-organizational feed validation falls outside \fname's trust perimeter (Table~\ref{tab:attack_paths}, AP-4), and the framework lacks a standardized agentic authorization mechanism for autonomous credential lifecycle management given OAuth~2.1's unsuitability for long-running agent sessions~\cite{owasp2025agentic}. Our evaluation relies on structural analysis rather than adversarial simulation, and runtime overhead remains unmeasured; the field also lacks standardized datasets and benchmarks for multi-agent security evaluation. Priority directions include formal verification of consensus integrity under Byzantine conditions, empirical latency profiling under operational SOC workloads, multi-agent security benchmarks, and extension to federated topologies.

\section*{Conclusion}\label{section:conclusion}
    We presented \fname, a security framework for MAS grounded in the observation that documented attack vectors converge on two integration surfaces: tool orchestration and memory management. Five defensive principles, aligned with established compliance standards, provide full threat coverage across component, coordination, and protocol layers, and are instantiated in a CyberOps SOAR architecture that embeds security as an architectural constraint within the SOC lifecycle. Evaluation through coverage analysis, attack path tracing, and trust boundary assessment demonstrates a minimum of 72\% reduction in exploitable boundaries. Open challenges remain: single-point-of-failure resilience, consensus integrity under adversarial conditions, and validation latency in time-critical response. Empirical profiling in production enterprise environments with standardized dataset covering diverse use-cases and benchmark development are essential directions for future research.

\section*{Acknowledgement}
    This research was carried out in the PATENT Lab within the Department of Computer Science at The University of Alabama. The opinions and conclusions expressed in this paper are those of the authors and do not necessarily represent the official policies or positions of their affiliated institutions. The authors would also like to acknowledge the use of Flaticon in the diagrams.

\bibliographystyle{ACM-Reference-Format}
\bibliography{references}

\newpage
\appendix
\section{Appendix - List of Acronyms}
\small
\begin{tabular}{@{}ll@{}}
    A2A      & Agent-to-Agent Protocol \\
    AAR      & After Action Report \\
    ABAC     & Attribute-Based Access Control \\
    AI       & Artificial Intelligence \\
    API      & Application Programming Interface \\
    APM      & Access Policy Manager \\
    AV       & Antivirus  \\
    BCP      & Business Continuity Plan \\
    CMDB     & Configuration Management Database \\
    CMS      & Configuration Management System \\
    CoT      & Chain-of-Thought \\
    CTI      & Cyber Threat Intelligence \\
    CVE      & Common Vulnerabilities and Exposures \\
    CyberOps & Cybersecurity Operations \\
    EDR      & Endpoint Detection and Response \\
    EPP      & Endpoint Protection Platforms \\
    EU       & European Union \\
    GDPR     & General Data Protection Regulation \\
    GraphQL  & Graph Query Language \\
    GRC      & Governance, Risk, and Compliance \\
    I/O      & Input/Output \\
    IAM      & Identity and Access Management \\
    IDS      & Intrusion Detection System \\
    IoC      & Indicator of Compromise \\
    IoT      & Internet of Things \\
    IPS      & Intrusion Prevention System \\
    ISAC     & Information Sharing and Analysis Center \\
    ISO      & International Organization for Standardization \\
    ITSM     & Information Technology Service Management \\
    JSON     & JavaScript Object Notation \\
    LLM      & Large Language Model \\
    MAS      & Multi-Agent System \\
    MCP      & Model Context Protocol \\
    MISP     & Malware Information Sharing Platform \\
    ML       & Machine Learning \\
    MTTD     & Mean Time to Detect \\
    MTTR     & Mean Time to Respond \\
    NDR      & Network Detection and Response \\
    NIST     & National Institute of Standards and Technology \\
    NVD      & National Vulnerability Database \\
    OAuth    & Open Authorization \\
    OSINT    & Open-Source Intelligence \\
    OWASP    & Open Worldwide Application Security Project \\
    PAM      & Privileged Access Management \\
    PFI      & Prompt Flow Integrity \\
    RAG      & Retrieval-Augmented Generation \\
    RBAC     & Role-Based Access Control \\
    RCA      & Root Cause Analysis \\
    REST     & Representational State Transfer \\
    RPC      & Remote Procedure Call \\
    SIEM     & Security Information and Event Management \\
    SOAP     & Simple Object Access Protocol \\
    SOAR     & Security Orchestration, Automation, and Response \\
    SOC      & Security Operations Center \\
    SP       & Special Publication (NIST) \\
    SQL      & Structured Query Language \\
    TTP      & Tactics, Techniques, and Procedures \\
    UEBA     & User and Entity Behavior Analytics \\
    XSS      & Cross-Site Scripting \\
\end{tabular}

\clearpage
\onecolumn
\section{Appendix -- Literature Foundation}\label{appendix:literature}
    
    The following table summarizes the primary literature on which 
    \fname's attack surface decomposition and defensive framework 
    are built. For each work, we note its contribution to our 
    analysis, its scope limitation, and the specific takeaway that 
    informs our framework design.
    \vspace{1em}
    
    \scriptsize
    \renewcommand{\arraystretch}{1.1}
    \begin{longtable}{p{2.2cm} p{3.8cm} p{3.2cm} p{3.2cm} p{3cm}}
    \caption{Literature Foundation: Attack Surface Identification, 
    Defensive Mechanisms, and CyberOps Domain}
    \label{tab:literature_summary} \\
    
    \toprule
    \rowcolor{gray!30}
    \textbf{Reference} & \textbf{Contribution} & \textbf{Scope / Limitation} & \textbf{\fname Takeaway} & \textbf{Maps To} \\
    \midrule
    \endfirsthead
    
    \toprule
    \rowcolor{gray!30}
    \textbf{Reference} & \textbf{Contribution} & \textbf{Scope / Limitation} & \textbf{\fname Takeaway} & \textbf{Maps To} \\
    \midrule
    \endhead
    
    \midrule
    \multicolumn{5}{r}{\footnotesize\textit{Continued on next page}} \\
    \bottomrule
    \endfoot
    
    \bottomrule
    \endlastfoot
    
    \rowcolor{gray!10}
    \multicolumn{5}{l}{\textit{Component-Level Attack Surface (\S\ref{subsection:agenticai_components})}} \\
    
    Greshake et al.~\cite{greshake2023not}
    & Indirect prompt injection via retrieved external data; validated on Bing Chat and GitHub Copilot.
    & Single-agent pipeline; no multi-agent propagation.
    & Blurs data/instruction boundaries, enabling tool redirection and credential exfiltration; motivates write-boundary filtering.
    & Table~\ref{tab:attack_surface_mapping}: Context contamination \\
    
    Perez et al.~\cite{perez2022red}
    & Automated red-teaming to elicit harmful LLM outputs via adversarial prompts.
    & Model-level robustness only; no integration architecture.
    & Cognitive manipulation scoped out; integration-layer defenses address downstream effects.
    & Model-level scope exclusion (\S\ref{subsection:agenticai_components}) \\
    
    Carlini et al.~\cite{carlini2021extracting}
    & Training data extraction from LLMs via targeted querying.
    & Single-model privacy risk; no shared memory across agents.
    & Motivates memory isolation to prevent cross-agent data leakage.
    & P5: Access-Controlled Isolation \\
    
    Fang et al.~\cite{fang2024llm}
    & LLM agents autonomously hacking websites via SQLi and XSS; GPT-4 achieves 73.3\% success rate.
    & Offensive study; no defenses or multi-agent tool misuse.
    & Autonomous offensive tool invocation motivates capability scoping and least-privilege enforcement.
    & P2: Capability Scoping \\
    
    \midrule
    \rowcolor{gray!10}
    \multicolumn{5}{l}{\textit{Coordination-Level Attack Surface (\S\ref{subsection:aigent_types})}} \\
    
    Chen et al.~\cite{chen2024blockagents}
    & BlockAgents: Byzantine-robust MAS using blockchain proof-of-thought consensus against collusion and poisoning.
    & Consensus-layer only; no tool orchestration or memory isolation.
    & Byzantine collusion in MAS motivates consensus validation and verified execution.
    & Table~\ref{tab:attack_surface_mapping}: Lateral compromise \\
    
    Douceur~\cite{douceur2002sybil}
    & Formalizes Sybil attacks on decentralized systems via forged identities.
    & Classic distributed systems; not agent-specific.
    & Sybil attack via forged identities informs identity-binding and authentication at agent interfaces.
    & P1: Authorized Interface \\
    
    Agrawal et al.~\cite{agrawal2025evaluating}
    & Emergent collusion in frontier LLMs without explicit prompting.
    & Evaluative; no defensive mechanism proposed.
    & Collusion as coordination-layer threat motivates capability auditing and consensus validation.
    & Table~\ref{tab:attack_surface_mapping}: Covert coordination \\
    
    Mathew et al.~\cite{mathew2025hidden}
    & Steganographic collusion emerges unintentionally in LLMs via misspecified reward incentives, bypassing oversight.
    & Detection-focused; limited encoding schemes.
    & Emergent covert channels bypass traditional monitoring; motivates memory isolation and output filtering.
    & Table~\ref{tab:attack_surface_mapping}: Covert coordination \\
    
    \midrule
    \rowcolor{gray!10}
    \multicolumn{5}{l}{\textit{Protocol-Level Attack Surface (\S\ref{subsection:agenticai_standards})}} \\
    
    Triedman et al.~\cite{triedman2025multi}
    & MAS control-flow hijacking via inter-agent trust exploitation; validated on AutoGen, CrewAI, MetaGPT.
    & MAS-specific; no memory vectors or defensive mechanisms.
    & Control-flow hijacking via confused deputy exploitation motivates authorized interfaces and verified execution.
    & P1, P3; Table~\ref{tab:attack_surface_mapping}: Confused deputy \\
    
    De Pasquale et al.~\cite{de2025open}
    & Taxonomy of multi-agent threats: secret collusion, cascade attacks, and exploitation.
    & Broad taxonomy; no unified defensive model or validation.
    & Multi-agent threat taxonomy motivates defense-in-depth across coordination and protocol surfaces.
    & Table~\ref{tab:attack_surface_mapping}: Protocol-level vectors \\
    
    Shi et al.~\cite{shi2025towards}
    & Survey of GUI agent security; identifies adversarial input vulnerabilities and proposes input validation and sandboxing.
    & GUI-agent specific; no MAS coordination or memory vectors.
    & Adversarial input at perception boundary motivates write-boundary filtering and authorized interface enforcement.
    & P1: Authorized Interface \\
    
    He et al.~\cite{he2025red}
    & Agent-in-the-Middle (AiTM) attack intercepts and manipulates inter-agent communications; validated on MetaGPT and ChatDev.
    & Communication-focused; limited defenses, no memory or tool vectors.
    & Inter-agent message manipulation motivates verified execution and authorized interface enforcement.
    & Table~\ref{tab:attack_surface_mapping}: Message manipulation \\
    
    OWASP~\cite{owasp2025agentic}
    & Documents top agentic AI risks including identity/privilege abuse and excessive agency; notes OAuth~2.1 insufficiency for long-running sessions.
    & Standards gap analysis; no alternative authorization framework.
    & Identity and privilege abuse motivate agentic-specific interface authorization and capability scoping.
    & P1: Authorized Interface \\

    Anbiaee et al.~\cite{anbiaee2026security}
    & Comparative threat modeling of MCP, A2A, Agora, and 
    ANP; identifies twelve protocol-level risks across 
    lifecycle phases.
    & Risk taxonomy only; no defensive architecture or 
    empirical validation.
    & Cross-protocol risk fragmentation validates 
    protocol-agnostic defensive design in \fname.
    & Table~\ref{tab:attack_surface_mapping}: Protocol-level 
    vectors \\
    
    \midrule
    \rowcolor{gray!10}
    \multicolumn{5}{l}{\textit{Tool Integration Defenses (\S\ref{subsubsec:tool_authorized_interface}--\S\ref{subsubsec:tool_verified_actions})}} \\
    
    Narajala \& Sridhar~\cite{narajala2025securing}
    & Tool squatting threat and Zero Trust Tool Registry with admin-verified catalogs and JIT credential provisioning.
    & Architectural proposal; no empirical validation.
    & Tool squatting motivates signed manifests and admin-approved catalogs in P1.
    & P1: Authorized Interface \\
    
    Narajala \& Sridhar~\cite{narajala2025enterprise}
    & MCP security framework using Zero Trust, addressing tool poisoning, exfiltration, and C2 via cryptographic signing and strict I/O validation.
    & MCP-specific; implementation complexity limits adoption.
    & MCP threat landscape and Zero Trust mitigations motivate cryptographic provenance in P1.
    & P1: Authorized Interface \\
    
    Syros et al.~\cite{syros2025saga}
    & SAGA: cryptographic identity binding, user-controlled agent lifecycle, and token-based inter-agent communication; evaluated on eight attacker models.
    & Centralized architecture; no memory isolation.
    & Cryptographic identity binding and token-based access control complement registry-based discovery in P1.
    & P1: Authorized Interface \\
    
    Kim et al.~\cite{kim2025prompt}
    & PFI: privilege escalation prevention via agent isolation, untrusted data guardrails, and least-privilege enforcement; evaluated on AgentDojo and AgentBench.
    & Single-agent; limited to prompt and data injection vectors.
    & PFI's least-privilege enforcement and isolation guardrails directly inform P2 capability scoping.
    & P2: Capability Scoping \\
    
    Krawiecka et al.~\cite{krawiecka2025extending}
    & Extends OWASP MAS Threat Modeling Guide; identifies unsafe delegation escalation and emergent covert coordination.
    & Taxonomy extension; no empirical validation or enforcement mechanisms.
    & Unsafe delegation escalation motivates capability scoping and continuous privilege auditing.
    & P2: Capability Scoping \\
    
    De Pasquale et al.~\cite{de2024chainreactor}
    & ChainReactor: automated privilege escalation chain discovery; finds 20 novel chains in real-world AWS EC2 and Digital Ocean instances.
    & Offensive tool; no initial infiltration or exploit code generation.
    & Automated privilege escalation discovery motivates capability scoping and least-privilege enforcement.
    & P2: Capability Scoping \\
    
    Zou et al.~\cite{zou2025blocka2a}
    & BlockA2A: three-layer trust framework (Identity, Ledger, Smart Contract) with decentralized identity, immutable provenance, and programmable governance.
    & Blockchain overhead; on-chain operations require batching.
    & Decentralized identity binding and immutable audit trails inform verified execution and non-repudiation in P3.
    & P3: Verified Execution \\
    
    \midrule
    \rowcolor{gray!10}
    \multicolumn{5}{l}{\textit{Memory Management Defenses (\S\ref{subsubsec:memory_integrity}--\S\ref{subsubsec:memory_authentication})}} \\
    
    Wei et al.~\cite{wei2025memguard}
    & A-MemGuard: consensus-based validation and dual-memory structure to detect memory injection attacks in LLM agents.
    & Primarily single-agent; multi-agent scalability not fully explored.
    & Consensus-based memory validation and dual-memory structure inform write-boundary filtering and poisoning detection in P4.
    & P4: Memory Integrity \\
    
    Mao et al.~\cite{mao2025agentsafe}
    & AgentSafe: ThreatSieve for permission-controlled communication and HierarCache for hierarchical memory isolation; evaluated across multiple topologies.
    & Simulated environments; no real-world deployment validation.
    & Hierarchical memory isolation and permission-controlled communication inform P4 integrity and P5 isolation.
    & P4, P5 \\
    
    Xiang et al.~\cite{xiang2024certifiably}
    & RobustRAG: certifiably robust RAG via isolate-then-aggregate with keyword and decoding-based aggregation against retrieval corruption.
    & Bounded adversary; limited to passage-level corruption and single-hop retrieval.
    & Isolate-then-aggregate with certifiable robustness adopted for consensus-validated retrieval in P4.
    & P4: Memory Integrity \\
    
    Chen et al.~\cite{chen2024agentpoison}
    & AGENTPOISON: backdoor attack optimizing stealthy triggers to poison RAG-based agent memory while preserving benign performance.
    & Attack-focused; limited defense proposals.
    & Backdoor trigger poisoning of RAG memory motivates write-boundary filtering and provenance tracking in P4.
    & P4: Memory Integrity \\
    
    Wu et al.~\cite{wu2025memory}
    & Survey of memory mechanisms, architectures, and consistency challenges in LLM-MAS covering topology patterns and synchronization protocols.
    & Survey only; no persistent memory framework recommendations.
    & Memory consistency protocols across shared/local topologies motivate serialization and versioning in P4.
    & P4: Memory Integrity \\
    
    Wang et al.~\cite{wang2025unveiling}
    & MEXTRA: black-box memory extraction using locator-aligner prompts to extract private queries from LLM agent memory.
    & Single-agent; no multi-agent leakage or session control.
    & Black-box memory extraction motivates strict access compartmentalization and memory sanitization in P5.
    & P5: Access-Controlled Isolation \\
    
    Yagoubi et al.~\cite{yagoubi2026agentleak}
    & AgentLeak: full-stack benchmark evaluating privacy leakage across MAS channels via 1,000 scenarios and 32-class taxonomy.
    & Limited scenario coverage, language, and topology scope.
    & Internal channel leakage and default-open sharing motivate memory access controls and default-deny isolation in P5.
    & P5: Access-Controlled Isolation \\
    
    Rezazadeh et al.~\cite{rezazadeh2025collaborative}
    & Collaborative Memory: multi-user MAS with dynamic bipartite access graphs, two-tier private/shared memory, and granular read/write policies.
    & Controlled environments; probabilistic LLM behavior limits deterministic guarantees.
    & Dynamic bipartite access graphs and tiered memory isolation inform access control in P5.
    & P5: Access-Controlled Isolation \\
    
    Yuen et al.~\cite{yuen2025intrinsic}
    & Intrinsic Memory Agents: structured agent-specific memory templates to maintain role-relevant context in heterogeneous MAS.
    & Manual template creation; limited task diversity validation.
    & Agent-specific structured memory isolation informs role-scoped boundaries and access containment in P5.
    & P5: Access-Controlled Isolation \\
    
    Srivastava et al.~\cite{srivastava2025memorygraft}
    & MemoryGraft: two-phase attack injecting poisoned experience records into agent memory, causing persistent behavioral drift via semantic retrieval.
    & Single-agent; limited to aggregate retrieval statistics.
    & Trigger-free persistent memory poisoning motivates cryptographic provenance and write-boundary consensus in P4.
    & P4: Memory Integrity \\
    
    \midrule
    \rowcolor{gray!10}
    \multicolumn{5}{l}{\textit{CyberOps Domain and Architecture (\S\ref{section:agentic_cyber_tools})}} \\
    
    Ismail et al.~\cite{ismail2025toward}
    & Hyper-automation SOAR using agentic LLMs with IVAM lifecycle and dynamic playbook generation; reduces a 38-step playbook to 10 actions on Wazuh SIEM.
    & Pilot deployment; limited scenarios and AI explainability not addressed.
    & Agentic SOAR with dynamic playbook adaptation and human-in-the-loop motivates agentic integration in SOC workflows.
    & \S4.1: Operational gaps \\
    
    Mitra et al.~\cite{mitra2024localintel}
    & LocalIntel: contextualized CTI reasoning with agentic autonomy.
    & Single-agent; no multi-agent coordination or memory isolation.
    & Validates demand for context-aware reasoning; gaps motivate full MAS.
    & \S\ref{subsubsection:analytical_reasoning}: Analytical Reasoning, \S\ref{subsubsection:research} Research \\
    
    Mitra et al.~\cite{mitra2025falconautonomouscyberthreat}
    & Falcon: autonomous policy generation with iterative hypothesis refinement.
    & Single-agent threat hunting; no tool/memory security model.
    & Adaptive scouting adopted; security gaps motivate our defensive overlay.
    & \S\ref{subsubsection:decision_making}: Decision Making, \S\ref{subsubsection:research} Adaptive Improvement  \\
    
    Challita et al.~\cite{challita2025redteamllmagenticaiframework}
    & RedTeamLLM: agentic offensive security with dynamic plan correction, memory management, and ReAct reasoning; outperforms PentestGPT on VULNHUB.
    & Proof-of-concept; memory management and plan correction not fully implemented.
    & Agentic offensive capabilities with adaptive planning validate the need for dynamic playbooks in CyberOps.
    & \S\ref{subsubsection:research}: Adaptive Improvement \\
    
    Tomašev et al.~\cite{tomavsev2026intelligent}
    & Intelligent delegation framework with requirements 
    for dynamic assessment, adaptive execution, structural 
    transparency, scalable coordination, and systemic 
    resilience; mapped to MCP, A2A, AP2, and UCP protocols.
    & Theoretical only; no empirical implementation or 
    validation.
    & Delegation requirements for trust calibration and 
    systemic resilience align with \fname's verified 
    execution design.
    & \S5: Architectural alignment \\
    
    Lupinacci~\cite{lupinacci2025dark}
    & Three attack techniques; inter-agent trust exploitation achieves 100\% success across all tested models, enabling computer takeover via prompt injection and RAG backdoors.
    & Synthetic environments; no defensive framework proposed.
    & Universal inter-agent trust exploitation and RAG backdoors validate tool and memory surface vulnerabilities across component and coordination layers.
    & Table~\ref{tab:attack_surface_mapping}: Coverage validation \\
    
    \end{longtable}

\scriptsize
\clearpage
\onecolumn
\section{Appendix: Trust Boundary Enumeration}\label{appendix:trust_boundaries}

    This appendix enumerates all 200 trust boundaries in a flat 
    (unrestricted) MAS deployment and identifies which are 
    \textit{eliminated} by \fname's phase-scoping and mediation 
    architecture, and which are \textit{retained} with active 
    verification. The enumeration supports the 72\% reduction 
    claim (200$\rightarrow$56) in 
    Table~\ref{tab:trust_boundaries}.
    
    \vspace{1em}
    \noindent\textbf{Architecture Parameters.}
    \begin{itemize}
        \item \textbf{Agents (4):} A1=Monitor, A2=Analyze, 
        A3=Admin, A4=Report
        \item \textbf{Tools (16):} 
        T1=UEBA Model, T2=IDS/IPS/CMDB, T3=EDR/NDR Sensor, 
        T4=ITSM/Ticketing (Monitor); 
        T5=Sandbox, T6=SIEM Search, T7=Code Analyzer (Analyze); 
        T8=IAM/PAM, T9=Firewall/APM, T10=Configuration Manager, 
        T11=EPP/AV, T12=ANSIBLE (Admin); 
        T13=Reporting Dashboard, T14=ISAC/MISP, T15=Editor \& 
        Test Suite, T16=GRC Mapper (Report)
        \item \textbf{Memory Stores (12):} 
        M1=Threat Repository, M2=Asset Inventory, M3=Policy 
        Store, M4=SIEM Data Lake, M5=Code Repository, 
        M6=Case Management, M7=CTI Knowledge Base, 
        M8=Playbook Repository, M9=Compliance Mappings, 
        M10=Detection Rules, M11=AAR Archive, M12=BCP/Risk 
        Registry
        \item \textbf{External Feeds (12):} 
        E1=Vendor Advisories, E2=ISAC Feeds, E3=MISP 
        Community, E4=CVE/NVD, E5=OSINT Sources, 
        E6=Regulatory Updates, E7=Threat Intel Subs., 
        E8=Vulnerability Scanners, E9=Partner SOC Feeds, 
        E10=Industry Benchmarks, E11=Patch Bulletins, 
        E12=Compliance Updates
    \end{itemize}

    \vspace{0.3em}
    \noindent\textbf{Phase-to-Tool Assignment (P2: Capability 
    Scoping).}
    A1$\rightarrow$\{T1--T4\},\quad 
    A2$\rightarrow$\{T5--T7\},\quad 
    A3$\rightarrow$\{T8--T12\},\quad 
    A4$\rightarrow$\{T13--T16\}
    
    \vspace{0.3em}
    \noindent\textbf{Phase-to-Memory Assignment (P5: Access 
    Control with Data Isolation).}
    A1$\rightarrow$\{M1, M4, M7, M10\},\quad 
    A2$\rightarrow$\{M1, M2, M6, M7\},\quad 
    A3$\rightarrow$\{M2, M3, M8, M12\},\quad 
    A4$\rightarrow$\{M8, M9, M10, M11\}
    
    \vspace{0.3em}
    \noindent\textbf{Legend.}
    $\times$\,=\,Retained (with active verification 
    mechanism);\quad 
    \checkmark\,=\,Eliminated by \fname.\vspace{1em}

    \vspace{1em}
    \noindent\textbf{Category 1: Agent $\rightarrow$ Tool Boundaries} \hfill \textit{64 total, 16 retained, 48 eliminated (75\%)}

    \noindent Phase-scoping (P2) restricts each agent to its assigned tools (Monitor: 4, Analyze: 3, Admin: 5, Report: 4). Retained boundaries are subject to signed manifest verification (P1) and consensus-validated execution (P3).\\

    \begin{longtable}{p{0.6cm} p{3.5cm} p{3.5cm} p{1cm} p{6cm}}
    \toprule
    \rowcolor{gray!30}
    \textbf{\#} & \textbf{Agent} & \textbf{Tool} & \textbf{Status} & \textbf{Mechanism} \\
    \midrule
    \endfirsthead
    \toprule
    \rowcolor{gray!30}
    \textbf{\#} & \textbf{Agent} & \textbf{Tool} & \textbf{Status} & \textbf{Mechanism} \\
    \midrule
    \endhead
    \midrule
    \multicolumn{5}{r}{\footnotesize\textit{Continued on next page}} \\
    \bottomrule
    \endfoot
    \bottomrule
    \endlastfoot
    
    \rowcolor{gray!10}
    \multicolumn{5}{l}{\textit{A1 (Monitor) $\rightarrow$ Tools [4 assigned: T1--T4]}} \\
        1  & A1-Monitor & T1-UEBA Model        & $\times$ & P1+P3: Signed manifest, Validation Loop \\
        2  & A1-Monitor & T2-IDS/IPS/CMDB      & $\times$ & P1+P3: Signed manifest, Validation Loop \\
        3  & A1-Monitor & T3-EDR/NDR Sensor     & $\times$ & P1+P3: Signed manifest, Validation Loop \\
        4  & A1-Monitor & T4-ITSM/Ticketing     & $\times$ & P1+P3: Signed manifest, Validation Loop \\
        5  & A1-Monitor & T5-Sandbox            & \checkmark   & P2: Not in Monitor manifest \\
        6  & A1-Monitor & T6-SIEM Search        & \checkmark   & P2: Not in Monitor manifest \\
        7  & A1-Monitor & T7-Code Analyzer      & \checkmark   & P2: Not in Monitor manifest \\
        8  & A1-Monitor & T8-IAM/PAM            & \checkmark   & P2: Not in Monitor manifest \\
        9  & A1-Monitor & T9-Firewall/APM       & \checkmark   & P2: Not in Monitor manifest \\
        10 & A1-Monitor & T10-Config.\ Manager  & \checkmark   & P2: Not in Monitor manifest \\
        11 & A1-Monitor & T11-EPP/AV            & \checkmark   & P2: Not in Monitor manifest \\
        12 & A1-Monitor & T12-ANSIBLE           & \checkmark   & P2: Not in Monitor manifest \\
        13 & A1-Monitor & T13-Report Dashboard  & \checkmark   & P2: Not in Monitor manifest \\
        14 & A1-Monitor & T14-ISAC/MISP         & \checkmark   & P2: Not in Monitor manifest \\
        15 & A1-Monitor & T15-Editor \& Test    & \checkmark   & P2: Not in Monitor manifest \\
        16 & A1-Monitor & T16-GRC Mapper        & \checkmark   & P2: Not in Monitor manifest \\
        \midrule
        
    \rowcolor{gray!10}
    \multicolumn{5}{l}{\textit{A2 (Analyze) $\rightarrow$ Tools [3 assigned: T5--T7]}} \\
        17 & A2-Analyze & T1-UEBA Model        & \checkmark   & P2: Not in Analyze manifest \\
        18 & A2-Analyze & T2-IDS/IPS/CMDB      & \checkmark   & P2: Not in Analyze manifest \\
        19 & A2-Analyze & T3-EDR/NDR Sensor     & \checkmark   & P2: Not in Analyze manifest \\
        20 & A2-Analyze & T4-ITSM/Ticketing     & \checkmark   & P2: Not in Analyze manifest \\
        21 & A2-Analyze & T5-Sandbox            & $\times$ & P1+P3: Signed manifest, RCA Loop \\
        22 & A2-Analyze & T6-SIEM Search        & $\times$ & P1+P3: Signed manifest, RCA Loop \\
        23 & A2-Analyze & T7-Code Analyzer      & $\times$ & P1+P3: Signed manifest, RCA Loop \\
        24 & A2-Analyze & T8-IAM/PAM            & \checkmark   & P2: Not in Analyze manifest \\
        25 & A2-Analyze & T9-Firewall/APM       & \checkmark   & P2: Not in Analyze manifest \\
        26 & A2-Analyze & T10-Config.\ Manager  & \checkmark   & P2: Not in Analyze manifest \\
        27 & A2-Analyze & T11-EPP/AV            & \checkmark   & P2: Not in Analyze manifest \\
        28 & A2-Analyze & T12-ANSIBLE           & \checkmark   & P2: Not in Analyze manifest \\
        29 & A2-Analyze & T13-Report Dashboard  & \checkmark   & P2: Not in Analyze manifest \\
        30 & A2-Analyze & T14-ISAC/MISP         & \checkmark   & P2: Not in Analyze manifest \\
        31 & A2-Analyze & T15-Editor \& Test    & \checkmark   & P2: Not in Analyze manifest \\
        32 & A2-Analyze & T16-GRC Mapper        & \checkmark   & P2: Not in Analyze manifest \\
        \midrule
        
    \rowcolor{gray!10}
    \multicolumn{5}{l}{\textit{A3 (Admin) $\rightarrow$ Tools [5 assigned: T8--T12]}} \\
        33 & A3-Admin & T1-UEBA Model          & \checkmark   & P2: Not in Admin manifest \\
        34 & A3-Admin & T2-IDS/IPS/CMDB        & \checkmark   & P2: Not in Admin manifest \\
        35 & A3-Admin & T3-EDR/NDR Sensor       & \checkmark   & P2: Not in Admin manifest \\
        36 & A3-Admin & T4-ITSM/Ticketing       & \checkmark   & P2: Not in Admin manifest \\
        37 & A3-Admin & T5-Sandbox              & \checkmark   & P2: Not in Admin manifest \\
        38 & A3-Admin & T6-SIEM Search          & \checkmark   & P2: Not in Admin manifest \\
        39 & A3-Admin & T7-Code Analyzer        & \checkmark   & P2: Not in Admin manifest \\
        40 & A3-Admin & T8-IAM/PAM              & $\times$ & P1+P3: Signed manifest, Recovery Loop \\
        41 & A3-Admin & T9-Firewall/APM         & $\times$ & P1+P3: Signed manifest, Recovery Loop \\
        42 & A3-Admin & T10-Config.\ Manager    & $\times$ & P1+P3: Signed manifest, Recovery Loop \\
        43 & A3-Admin & T11-EPP/AV              & $\times$ & P1+P3: Signed manifest, Recovery Loop \\
        44 & A3-Admin & T12-ANSIBLE             & $\times$ & P1+P3: Signed manifest, Recovery Loop \\
        45 & A3-Admin & T13-Report Dashboard    & \checkmark   & P2: Not in Admin manifest \\
        46 & A3-Admin & T14-ISAC/MISP           & \checkmark   & P2: Not in Admin manifest \\
        47 & A3-Admin & T15-Editor \& Test      & \checkmark   & P2: Not in Admin manifest \\
        48 & A3-Admin & T16-GRC Mapper          & \checkmark   & P2: Not in Admin manifest \\
        \midrule
        
    \rowcolor{gray!10}
    \multicolumn{5}{l}{\textit{A4 (Report) $\rightarrow$ Tools [4 assigned: T13--T16]}} \\
        49 & A4-Report & T1-UEBA Model         & \checkmark   & P2: Not in Report manifest \\
        50 & A4-Report & T2-IDS/IPS/CMDB       & \checkmark   & P2: Not in Report manifest \\
        51 & A4-Report & T3-EDR/NDR Sensor      & \checkmark   & P2: Not in Report manifest \\
        52 & A4-Report & T4-ITSM/Ticketing      & \checkmark   & P2: Not in Report manifest \\
        53 & A4-Report & T5-Sandbox             & \checkmark   & P2: Not in Report manifest \\
        54 & A4-Report & T6-SIEM Search         & \checkmark   & P2: Not in Report manifest \\
        55 & A4-Report & T7-Code Analyzer       & \checkmark   & P2: Not in Report manifest \\
        56 & A4-Report & T8-IAM/PAM             & \checkmark   & P2: Not in Report manifest \\
        57 & A4-Report & T9-Firewall/APM        & \checkmark   & P2: Not in Report manifest \\
        58 & A4-Report & T10-Config.\ Manager   & \checkmark   & P2: Not in Report manifest \\
        59 & A4-Report & T11-EPP/AV             & \checkmark   & P2: Not in Report manifest \\
        60 & A4-Report & T12-ANSIBLE            & \checkmark   & P2: Not in Report manifest \\
        61 & A4-Report & T13-Report Dashboard   & $\times$ & P1+P3: Signed manifest, Improvement Loop \\
        62 & A4-Report & T14-ISAC/MISP          & $\times$ & P1+P3: Signed manifest, Improvement Loop \\
        63 & A4-Report & T15-Editor \& Test     & $\times$ & P1+P3: Signed manifest, Improvement Loop \\
        64 & A4-Report & T16-GRC Mapper         & $\times$ & P1+P3: Signed manifest, Improvement Loop \\
    
    \end{longtable}
    
    \vspace{0.3em}
    \noindent\textbf{Category 1 Summary:} 16 retained ($\times$), 48 eliminated (\checkmark). Reduction: 75\%.

    \vspace{1em}
    \noindent\textbf{Category 2: Agent $\rightarrow$ Memory Boundaries} \hfill \textit{48 total, 16 retained, 32 eliminated (67\%)}
    
    \noindent Phase-partitioned access (P5) restricts each agent to four designated memory stores. All retained boundaries are mediated by the Memory Management Agent with write-boundary filtering (P4).
    
    \begin{longtable}{p{0.6cm} p{3.5cm} p{3.5cm} p{1cm} p{6cm}}
    \toprule
    \rowcolor{gray!30}
    \textbf{\#} & \textbf{Agent} & \textbf{Memory Store} & \textbf{Status} & \textbf{Mechanism} \\
    \midrule
    \endfirsthead
    \toprule
    \rowcolor{gray!30}
    \textbf{\#} & \textbf{Agent} & \textbf{Memory Store} & \textbf{Status} & \textbf{Mechanism} \\
    \midrule
    \endhead
    \midrule
    \multicolumn{5}{r}{\footnotesize\textit{Continued on next page}} \\
    \bottomrule
    \endfoot
    \bottomrule
    \endlastfoot
    
    \rowcolor{gray!10}
    \multicolumn{5}{l}{\textit{A1 (Monitor) $\rightarrow$ Memory}} \\
    65 & A1-Monitor & M1-Threat Repository  & $\times$ & P4+P5: MMA-mediated, write filtering \\
    66 & A1-Monitor & M2-Asset Inventory    & \checkmark   & P5: Outside Monitor scope \\
    67 & A1-Monitor & M3-Policy Store       & \checkmark   & P5: Outside Monitor scope \\
    68 & A1-Monitor & M4-SIEM Data Lake     & $\times$ & P4+P5: MMA-mediated, write filtering \\
    69 & A1-Monitor & M5-Code Repository    & \checkmark   & P5: Outside Monitor scope \\
    70 & A1-Monitor & M6-Case Management    & \checkmark   & P5: Outside Monitor scope \\
    71 & A1-Monitor & M7-CTI Knowledge Base & $\times$ & P4+P5: MMA-mediated, read-only \\
    72 & A1-Monitor & M8-Playbook Repo      & \checkmark   & P5: Outside Monitor scope \\
    73 & A1-Monitor & M9-Compliance Map     & \checkmark   & P5: Outside Monitor scope \\
    74 & A1-Monitor & M10-Detection Rules   & $\times$ & P4+P5: MMA-mediated, read-only \\
    75 & A1-Monitor & M11-AAR Archive       & \checkmark   & P5: Outside Monitor scope \\
    76 & A1-Monitor & M12-BCP/Risk Reg.     & \checkmark   & P5: Outside Monitor scope \\
    \midrule
    
    \rowcolor{gray!10}
    \multicolumn{5}{l}{\textit{A2 (Analyze) $\rightarrow$ Memory}} \\
    77 & A2-Analyze & M1-Threat Repository  & $\times$ & P4+P5: MMA-mediated, write filtering \\
    78 & A2-Analyze & M2-Asset Inventory    & $\times$ & P4+P5: MMA-mediated, read-only \\
    79 & A2-Analyze & M3-Policy Store       & \checkmark   & P5: Outside Analyze scope \\
    80 & A2-Analyze & M4-SIEM Data Lake     & \checkmark   & P5: Outside Analyze scope \\
    81 & A2-Analyze & M5-Code Repository    & \checkmark   & P5: Outside Analyze scope \\
    82 & A2-Analyze & M6-Case Management    & $\times$ & P4+P5: MMA-mediated, write filtering \\
    83 & A2-Analyze & M7-CTI Knowledge Base & $\times$ & P4+P5: MMA-mediated, read-only \\
    84 & A2-Analyze & M8-Playbook Repo      & \checkmark   & P5: Outside Analyze scope \\
    85 & A2-Analyze & M9-Compliance Map     & \checkmark   & P5: Outside Analyze scope \\
    86 & A2-Analyze & M10-Detection Rules   & \checkmark   & P5: Outside Analyze scope \\
    87 & A2-Analyze & M11-AAR Archive       & \checkmark   & P5: Outside Analyze scope \\
    88 & A2-Analyze & M12-BCP/Risk Reg.     & \checkmark   & P5: Outside Analyze scope \\
    \midrule
    
    \rowcolor{gray!10}
    \multicolumn{5}{l}{\textit{A3 (Admin) $\rightarrow$ Memory}} \\
    89  & A3-Admin & M1-Threat Repository  & \checkmark   & P5: Outside Admin scope \\
    90  & A3-Admin & M2-Asset Inventory    & $\times$ & P4+P5: MMA-mediated, write filtering \\
    91  & A3-Admin & M3-Policy Store       & $\times$ & P4+P5: MMA-mediated, read-only \\
    92  & A3-Admin & M4-SIEM Data Lake     & \checkmark   & P5: Outside Admin scope \\
    93  & A3-Admin & M5-Code Repository    & \checkmark   & P5: Outside Admin scope \\
    94  & A3-Admin & M6-Case Management    & \checkmark   & P5: Outside Admin scope \\
    95  & A3-Admin & M7-CTI Knowledge Base & \checkmark   & P5: Outside Admin scope \\
    96  & A3-Admin & M8-Playbook Repo      & $\times$ & P4+P5: MMA-mediated, read-only \\
    97  & A3-Admin & M9-Compliance Map     & \checkmark   & P5: Outside Admin scope \\
    98  & A3-Admin & M10-Detection Rules   & \checkmark   & P5: Outside Admin scope \\
    99  & A3-Admin & M11-AAR Archive       & \checkmark   & P5: Outside Admin scope \\
    100 & A3-Admin & M12-BCP/Risk Reg.     & $\times$ & P4+P5: MMA-mediated, read-only \\
    \midrule
    
    \rowcolor{gray!10}
    \multicolumn{5}{l}{\textit{A4 (Report) $\rightarrow$ Memory}} \\
    101 & A4-Report & M1-Threat Repository  & \checkmark   & P5: Outside Report scope \\
    102 & A4-Report & M2-Asset Inventory    & \checkmark   & P5: Outside Report scope \\
    103 & A4-Report & M3-Policy Store       & \checkmark   & P5: Outside Report scope \\
    104 & A4-Report & M4-SIEM Data Lake     & \checkmark   & P5: Outside Report scope \\
    105 & A4-Report & M5-Code Repository    & \checkmark   & P5: Outside Report scope \\
    106 & A4-Report & M6-Case Management    & \checkmark   & P5: Outside Report scope \\
    107 & A4-Report & M7-CTI Knowledge Base & \checkmark   & P5: Outside Report scope \\
    108 & A4-Report & M8-Playbook Repo      & $\times$ & P4+P5: MMA-mediated, Improvement Loop \\
    109 & A4-Report & M9-Compliance Map     & $\times$ & P4+P5: MMA-mediated, Improvement Loop \\
    110 & A4-Report & M10-Detection Rules   & $\times$ & P4+P5: MMA-mediated, Improvement Loop \\
    111 & A4-Report & M11-AAR Archive       & $\times$ & P4+P5: MMA-mediated, Improvement Loop \\
    112 & A4-Report & M12-BCP/Risk Reg.     & \checkmark   & P5: Outside Report scope \\
    
    \end{longtable}
    
    \vspace{0.3em}
    \noindent\textbf{Category 2 Summary:} 16 retained ($\times$), 32 eliminated (\checkmark). Reduction: 67\%.
    
    \vspace{1em}
    \noindent\textbf{Category 3: Agent $\leftrightarrow$ Agent Boundaries} \hfill \textit{12 total, 4 retained, 8 eliminated (67\%)}
    
    \noindent In the flat baseline, every agent pair communicates bidirectionally (6 pairs $\times$ 2 directions = 12). \fname restricts inter-agent communication to sequential Host-mediated phase handoffs. Only four directional handoffs are retained, each traversing the Host as an intermediary trust checkpoint.
    
    \begin{longtable}{p{0.6cm} p{3.5cm} p{3.5cm} p{1cm} p{6cm}}
    \toprule
    \rowcolor{gray!30}
    \textbf{\#} & \textbf{Source} & \textbf{Destination} & \textbf{Status} & \textbf{Mechanism} \\
    \midrule
    \endfirsthead
    \toprule
    \rowcolor{gray!30}
    \textbf{\#} & \textbf{Source} & \textbf{Destination} & \textbf{Status} & \textbf{Mechanism} \\
    \midrule
    \endhead
    \bottomrule
    \endlastfoot
    
    113 & A1-Monitor & A2-Analyze  & $\times$ & Host-mediated, P1+P3 \\
    114 & A2-Analyze & A1-Monitor  & \checkmark   & No reverse handoff \\
    115 & A1-Monitor & A3-Admin    & \checkmark   & Non-adjacent phase \\
    116 & A3-Admin   & A1-Monitor  & \checkmark   & Non-adjacent phase \\
    117 & A1-Monitor & A4-Report   & \checkmark   & Non-adjacent phase \\
    118 & A4-Report  & A1-Monitor  & $\times$ & Host-mediated feedback loop \\
    119 & A2-Analyze & A3-Admin    & $\times$ & Host-mediated, P1+P3 \\
    120 & A3-Admin   & A2-Analyze  & \checkmark   & No reverse handoff \\
    121 & A2-Analyze & A4-Report   & \checkmark   & Non-adjacent phase \\
    122 & A4-Report  & A2-Analyze  & \checkmark   & Non-adjacent phase \\
    123 & A3-Admin   & A4-Report   & $\times$ & Host-mediated, P1+P3 \\
    124 & A4-Report  & A3-Admin    & \checkmark   & No reverse handoff \\
    
    \end{longtable}
    
    \vspace{0.3em}
    \noindent\textbf{Category 3 Summary:} 4 retained ($\times$), 8 eliminated (\checkmark). Reduction: 67\%.

    \vspace{1em}
    \noindent\textbf{Category 4: Tool Response $\rightarrow$ Agent Boundaries} \hfill \textit{64 total, 16 retained, 48 eliminated (75\%)}
    
    \noindent Tool responses are counted as separate boundaries because return payloads can carry injected content independent of the outbound request's authorization. In the flat baseline, any tool can return data to any agent. \fname restricts responses to the requesting agent's phase-scoped tools. Retained boundaries are subject to response verification via signed manifests (P1) and consensus validation (P3).
    
    \begin{longtable}{p{0.6cm} p{3.5cm} p{3.5cm} p{1cm} p{6cm}}
    \toprule
    \rowcolor{gray!30}
    \textbf{\#} & \textbf{Tool} & \textbf{Agent} & \textbf{Status} & \textbf{Mechanism} \\
    \midrule
    \endfirsthead
    \toprule
    \rowcolor{gray!30}
    \textbf{\#} & \textbf{Tool} & \textbf{Agent} & \textbf{Status} & \textbf{Mechanism} \\
    \midrule
    \endhead
    \midrule
    \multicolumn{5}{r}{\footnotesize\textit{Continued on next page}} \\
    \bottomrule
    \endfoot
    \bottomrule
    \endlastfoot
    
    \rowcolor{gray!10}
    \multicolumn{5}{l}{\textit{Monitor-Phase Tools (T1--T4) $\rightarrow$ Agents [4 tools]}} \\
    125 & T1-UEBA Model      & A1-Monitor & $\times$ & P1+P3: Signed response, Validation Loop \\
    126 & T1-UEBA Model      & A2-Analyze & \checkmark   & P2: Response blocked at Host \\
    127 & T1-UEBA Model      & A3-Admin   & \checkmark   & P2: Response blocked at Host \\
    128 & T1-UEBA Model      & A4-Report  & \checkmark   & P2: Response blocked at Host \\
    129 & T2-IDS/IPS/CMDB    & A1-Monitor & $\times$ & P1+P3: Signed response, Validation Loop \\
    130 & T2-IDS/IPS/CMDB    & A2-Analyze & \checkmark   & P2: Response blocked at Host \\
    131 & T2-IDS/IPS/CMDB    & A3-Admin   & \checkmark   & P2: Response blocked at Host \\
    132 & T2-IDS/IPS/CMDB    & A4-Report  & \checkmark   & P2: Response blocked at Host \\
    133 & T3-EDR/NDR Sensor  & A1-Monitor & $\times$ & P1+P3: Signed response, Validation Loop \\
    134 & T3-EDR/NDR Sensor  & A2-Analyze & \checkmark   & P2: Response blocked at Host \\
    135 & T3-EDR/NDR Sensor  & A3-Admin   & \checkmark   & P2: Response blocked at Host \\
    136 & T3-EDR/NDR Sensor  & A4-Report  & \checkmark   & P2: Response blocked at Host \\
    137 & T4-ITSM/Ticketing  & A1-Monitor & $\times$ & P1+P3: Signed response, Validation Loop \\
    138 & T4-ITSM/Ticketing  & A2-Analyze & \checkmark   & P2: Response blocked at Host \\
    139 & T4-ITSM/Ticketing  & A3-Admin   & \checkmark   & P2: Response blocked at Host \\
    140 & T4-ITSM/Ticketing  & A4-Report  & \checkmark   & P2: Response blocked at Host \\
    \midrule
    
    \rowcolor{gray!10}
    \multicolumn{5}{l}{\textit{Analyze-Phase Tools (T5--T7) $\rightarrow$ Agents [3 tools]}} \\
    141 & T5-Sandbox       & A1-Monitor & \checkmark   & P2: Response blocked at Host \\
    142 & T5-Sandbox       & A2-Analyze & $\times$ & P1+P3: Signed response, RCA Loop \\
    143 & T5-Sandbox       & A3-Admin   & \checkmark   & P2: Response blocked at Host \\
    144 & T5-Sandbox       & A4-Report  & \checkmark   & P2: Response blocked at Host \\
    145 & T6-SIEM Search   & A1-Monitor & \checkmark   & P2: Response blocked at Host \\
    146 & T6-SIEM Search   & A2-Analyze & $\times$ & P1+P3: Signed response, RCA Loop \\
    147 & T6-SIEM Search   & A3-Admin   & \checkmark   & P2: Response blocked at Host \\
    148 & T6-SIEM Search   & A4-Report  & \checkmark   & P2: Response blocked at Host \\
    149 & T7-Code Analyzer & A1-Monitor & \checkmark   & P2: Response blocked at Host \\
    150 & T7-Code Analyzer & A2-Analyze & $\times$ & P1+P3: Signed response, RCA Loop \\
    151 & T7-Code Analyzer & A3-Admin   & \checkmark   & P2: Response blocked at Host \\
    152 & T7-Code Analyzer & A4-Report  & \checkmark   & P2: Response blocked at Host \\
    \midrule
    
    \rowcolor{gray!10}
    \multicolumn{5}{l}{\textit{Admin-Phase Tools (T8--T12) $\rightarrow$ Agents [5 tools]}} \\
    153 & T8-IAM/PAM          & A1-Monitor & \checkmark   & P2: Response blocked at Host \\
    154 & T8-IAM/PAM          & A2-Analyze & \checkmark   & P2: Response blocked at Host \\
    155 & T8-IAM/PAM          & A3-Admin   & $\times$ & P1+P3: Signed response, Recovery Loop \\
    156 & T8-IAM/PAM          & A4-Report  & \checkmark   & P2: Response blocked at Host \\
    157 & T9-Firewall/APM     & A1-Monitor & \checkmark   & P2: Response blocked at Host \\
    158 & T9-Firewall/APM     & A2-Analyze & \checkmark   & P2: Response blocked at Host \\
    159 & T9-Firewall/APM     & A3-Admin   & $\times$ & P1+P3: Signed response, Recovery Loop \\
    160 & T9-Firewall/APM     & A4-Report  & \checkmark   & P2: Response blocked at Host \\
    161 & T10-Config.\ Manager & A1-Monitor & \checkmark  & P2: Response blocked at Host \\
    162 & T10-Config.\ Manager & A2-Analyze & \checkmark  & P2: Response blocked at Host \\
    163 & T10-Config.\ Manager & A3-Admin   & $\times$ & P1+P3: Signed response, Recovery Loop \\
    164 & T10-Config.\ Manager & A4-Report  & \checkmark  & P2: Response blocked at Host \\
    165 & T11-EPP/AV          & A1-Monitor & \checkmark   & P2: Response blocked at Host \\
    166 & T11-EPP/AV          & A2-Analyze & \checkmark   & P2: Response blocked at Host \\
    167 & T11-EPP/AV          & A3-Admin   & $\times$ & P1+P3: Signed response, Recovery Loop \\
    168 & T11-EPP/AV          & A4-Report  & \checkmark   & P2: Response blocked at Host \\
    169 & T12-ANSIBLE         & A1-Monitor & \checkmark   & P2: Response blocked at Host \\
    170 & T12-ANSIBLE         & A2-Analyze & \checkmark   & P2: Response blocked at Host \\
    171 & T12-ANSIBLE         & A3-Admin   & $\times$ & P1+P3: Signed response, Recovery Loop \\
    172 & T12-ANSIBLE         & A4-Report  & \checkmark   & P2: Response blocked at Host \\
    \midrule
    
    \rowcolor{gray!10}
    \multicolumn{5}{l}{\textit{Report-Phase Tools (T13--T16) $\rightarrow$ Agents [4 tools]}} \\
    173 & T13-Report Dashboard & A1-Monitor & \checkmark   & P2: Response blocked at Host \\
    174 & T13-Report Dashboard & A2-Analyze & \checkmark   & P2: Response blocked at Host \\
    175 & T13-Report Dashboard & A3-Admin   & \checkmark   & P2: Response blocked at Host \\
    176 & T13-Report Dashboard & A4-Report  & $\times$ & P1+P3: Signed response, Improvement Loop \\
    177 & T14-ISAC/MISP       & A1-Monitor & \checkmark   & P2: Response blocked at Host \\
    178 & T14-ISAC/MISP       & A2-Analyze & \checkmark   & P2: Response blocked at Host \\
    179 & T14-ISAC/MISP       & A3-Admin   & \checkmark   & P2: Response blocked at Host \\
    180 & T14-ISAC/MISP       & A4-Report  & $\times$ & P1+P3: Signed response, Improvement Loop \\
    181 & T15-Editor \& Test  & A1-Monitor & \checkmark   & P2: Response blocked at Host \\
    182 & T15-Editor \& Test  & A2-Analyze & \checkmark   & P2: Response blocked at Host \\
    183 & T15-Editor \& Test  & A3-Admin   & \checkmark   & P2: Response blocked at Host \\
    184 & T15-Editor \& Test  & A4-Report  & $\times$ & P1+P3: Signed response, Improvement Loop \\
    185 & T16-GRC Mapper      & A1-Monitor & \checkmark   & P2: Response blocked at Host \\
    186 & T16-GRC Mapper      & A2-Analyze & \checkmark   & P2: Response blocked at Host \\
    187 & T16-GRC Mapper      & A3-Admin   & \checkmark   & P2: Response blocked at Host \\
    188 & T16-GRC Mapper      & A4-Report  & $\times$ & P1+P3: Signed response, Improvement Loop \\
    
    \end{longtable}
    
    \vspace{0.3em}
    \noindent\textbf{Category 4 Summary:} 16 retained ($\times$), 48 eliminated (\checkmark). Reduction: 75\%.
    
    \vspace{1em}
    \noindent\textbf{Category 5: External Feed $\rightarrow$ Memory Boundaries} \hfill \textit{12 total, 4 retained, 8 eliminated (67\%)}
    
    \noindent In the flat baseline, each external feed writes directly to a corresponding memory store without mediation. \fname routes all external ingestion through the Memory Management Agent (MMA), which applies write-boundary filtering (P4) and restricts feeds to designated stores (P5). Only four feed-to-store paths are retained.
    
    \begin{longtable}{p{0.6cm} p{3.5cm} p{3.5cm} p{1cm} p{6cm}}
    \toprule
    \rowcolor{gray!30}
    \textbf{\#} & \textbf{External Feed} & \textbf{Memory Store} & \textbf{Status} & \textbf{Mechanism} \\
    \midrule
    \endfirsthead
    \toprule
    \rowcolor{gray!30}
    \textbf{\#} & \textbf{External Feed} & \textbf{Memory Store} & \textbf{Status} & \textbf{Mechanism} \\
    \midrule
    \endhead
    \bottomrule
    \endlastfoot
    
    189 & E1-Vendor Advisories       & M7-CTI Knowledge Base & $\times$ & P4+P5: MMA write filtering \\
    190 & E2-ISAC Feeds              & M7-CTI Knowledge Base & \checkmark   & P5: Consolidated via E1 path \\
    191 & E3-MISP Community          & M1-Threat Repository  & $\times$ & P4+P5: MMA write filtering \\
    192 & E4-CVE/NVD                 & M1-Threat Repository  & \checkmark   & P5: Consolidated via E3 path \\
    193 & E5-OSINT Sources           & M7-CTI Knowledge Base & \checkmark   & P5: Consolidated via E1 path \\
    194 & E6-Regulatory Updates      & M9-Compliance Map     & $\times$ & P4+P5: MMA write filtering \\
    195 & E7-Threat Intel Subs.      & M1-Threat Repository  & \checkmark   & P5: Consolidated via E3 path \\
    196 & E8-Vulnerability Scanners  & M4-SIEM Data Lake     & \checkmark   & P5: Ingested via Monitor tools \\
    197 & E9-Partner SOC Feeds       & M7-CTI Knowledge Base & \checkmark   & P5: Consolidated via E1 path \\
    198 & E10-Industry Benchmarks    & M9-Compliance Map     & \checkmark   & P5: Consolidated via E6 path \\
    199 & E11-Patch Bulletins        & M10-Detection Rules   & $\times$ & P4+P5: MMA write filtering \\
    200 & E12-Compliance Updates     & M9-Compliance Map     & \checkmark   & P5: Consolidated via E6 path \\
    
    \end{longtable}
    
    \vspace{0.3em}
    \noindent\textbf{Category 5 Summary:} 4 retained ($\times$), 8 eliminated (\checkmark). Reduction: 67\%.

    \vspace{1em}
    \begin{center}
    \begin{tcolorbox}[enhanced, colback=white, colframe=gray!75!black, 
        colbacktitle=gray!80!black, title=Trust Boundary Verification Summary, 
        left=2mm, right=2mm, boxrule=0.75pt, top=2mm, bottom=2mm, width=0.58\textwidth]
        \centering
        \scriptsize
        \begin{tabular}{p{4.2cm}rrr}
        \textbf{Category} & \textbf{Flat} & \textbf{\fname} & \textbf{Elim.} \\
        \midrule
        Cat.~1: Agent $\rightarrow$ Tool (\#1--64)           & 64  & 16 & 48 \\
        Cat.~2: Agent $\rightarrow$ Memory (\#65--112)       & 48  & 16 & 32 \\
        Cat.~3: Agent $\leftrightarrow$ Agent (\#113--124)   & 12  & 4  & 8 \\
        Cat.~4: Tool Resp. $\rightarrow$ Agent (\#125--188)  & 64  & 16 & 48 \\
        Cat.~5: Ext.~Feed $\rightarrow$ Memory (\#189--200)  & 12  & 4  & 8 \\
        \midrule
        \textbf{Total}                                        & \textbf{200} & \textbf{56} & \textbf{144} \\
        \end{tabular}
        
        \vspace{0.5em}
        \footnotesize
        \noindent\textbf{Reduction: 144/200 = 72\%.} All 56 
        retained boundaries are subject to at least one active 
        verification mechanism (P1, P3, P4, or P5).
    \end{tcolorbox}
    \end{center}










\end{document}